\documentclass[twocolumn]{aastex631}

\usepackage{savesym}
\savesymbol{tablenum}
\usepackage{siunitx}
\restoresymbol{SIX}{tablenum}

\usepackage[version=4]{mhchem}
\usepackage{graphicx,epsf}
\usepackage{subfigure}
\usepackage{graphicx}	% Including figure files
\usepackage{amsmath}	% Advanced maths commands
\usepackage{amssymb}	% Extra maths symbols
\usepackage{newtxtext,newtxmath}
\usepackage{siunitx}
\usepackage{footnote}

\newcommand{\Msun}{{\rm M}_\odot}

\newcommand{\kms}{\textrm{km}\,\textrm{s}^{-1}}

\def\arcsec{\hbox{$^{\prime\prime}$}}

\def\sn{{SN~2023ixf}}
\newcommand{\mdot}{M$_{\odot}$~yr$^{-1}$}

\DeclareRobustCommand{\ion}[2]{\relax\ifmmode\ifx\testbx\f@series{\mathbf{#1\,\mathsc{#2}}}\else{\mathrm{#1\,\mathsc{#2}}}\fi\else\textup{#1\,{\mdseries\textsc{#2}}}\fi}

\newcommand{\code}[1]{\texttt{#1}}

\def\cmfgen{{\code{CMFGEN}}}

\usepackage[scale=0.94]{tgbonum}
\usepackage[mode=text]{siunitx}

\makeatletter
\DeclareTextCompositeCommand{\r}{OT1}{A}{%
  \leavevmode\vbox{%
    \offinterlineskip
    \ialign{\hfil##\hfil\cr\char23\cr\noalign{\kern-1.15ex}A\cr}%
  }%
}
\makeatother

\shorttitle{Shock Power in Supernova 2023ixf}
\shortauthors{Jacobson-Gal\'an et al.}

\begin{document}

\title{A Panchromatic View of Late-time Shock Power in the Type II Supernova 2023ixf}

\correspondingauthor{Wynn Jacobson-Gal\'{a}n (he, him, his)}
\email{wynnjg@caltech.edu}

\author[0000-0002-3934-2644]{W.~V.~Jacobson-Gal\'{a}n}
\altaffiliation{NASA Hubble Fellow}
\affiliation{Cahill Center for Astrophysics, California Institute of Technology, MC 249-17, 1216 E California Boulevard, Pasadena, CA, 91125, USA}

\author[0000-0003-0599-8407]{L.~Dessart}
\affil{Institut d’Astrophysique de Paris, CNRS-Sorbonne Université, 98 bis boulevard Arago, F-75014 Paris, France}

\author[0000-0002-5740-7747]{C.~D.~Kilpatrick}
\affil{Center for Interdisciplinary Exploration and Research in Astrophysics (CIERA), Northwestern University, Evanston, IL 60202, USA}

\author{P.~J.~Patel}
\affil{Department of Astronomy and Astrophysics, University of Chicago, Eckhardt, 5640 South Ellis Avenue, Chicago, IL 60637, USA}

\author[0000-0002-4449-9152]{K.~Auchettl}
\affil{Department of Astronomy and Astrophysics, University of California, Santa Cruz, CA 95064, USA}
\affil{OzGrav, School of Physics, The University of Melbourne, VIC 3010, Australia}

\author[0000-0002-1481-4676]{S.~Tinyanont}
\affil{National Astronomical Research Institute of Thailand, 260 Moo 4, Donkaew, Maerim, Chiang Mai, 50180, Thailand}

\author[0000-0003-4768-7586]{R.~Margutti}
\affil{Department of Astronomy, University of California, Berkeley, CA 94720-3411, USA}
\affil{Department of Physics, University of California, Berkeley, CA 94720, USA}

\author[0000-0002-4661-7001]{V.~V.~Dwarkadas}
\affiliation{Department of Astronomy and Astrophysics, University of Chicago, Eckhardt, 5640 South Ellis Avenue, Chicago, IL 60637, USA}

\author[0000-0002-4924-444X]{K.~A.~Bostroem}
\affil{Steward Observatory, University of Arizona, 933 North Cherry Avenue, Tucson, AZ 85721-0065, USA}
\altaffiliation{LSST-DA Catalyst Fellow}

\author[0000-0002-7706-5668]{R.~Chornock}
\affil{Department of Astronomy, University of California, Berkeley, CA 94720-3411, USA}

\author[0000-0002-2445-5275]{R.~J.~Foley}
\affiliation{Department of Astronomy and Astrophysics, University of California, Santa Cruz, CA 95064, USA}

\author[0009-0000-7177-9697]{H.~Abunemeh}
\affiliation{Department of Astronomy, University of Illinois at Urbana-Champaign, 1002 W. Green St., IL 61801, USA}

\author[0000-0002-2184-6430]{T.~Ahumada}
\affiliation{Cahill Center for Astrophysics, California Institute of Technology, MC 249-17, 1216 E California Boulevard, Pasadena, CA, 91125, USA}

\author[0000-0002-6688-3307]{P.~Arunachalam}
\affiliation{Department of Astronomy and Astrophysics, University of California, Santa Cruz, CA 95064, USA}

\author[0000-0003-0416-9818]{M.~J.~Bustamante-Rosell}
\affiliation{Department of Life and Physical Sciences, Fisk University, 1000 17th Avenue N., Nashville, TN 37208, USA}
\affil{Department of Physics and Astronomy, Vanderbilt University, 6301 Stevenson Center Lane, Nashville, TN 37235, USA}

\author[0000-0003-4263-2228]{D.~A.~Coulter}
\affil{Space Telescope Science Institute, Baltimore, MD 21218, USA}

\author[0000-0002-8526-3963]{C.~Gall}
\affil{DARK, Niels Bohr Institute, University of Copenhagen, Jagtvej 128, 2200 Copenhagen, Denmark}

\author[0000-0003-1015-5367]{H.~Gao}
\affiliation{Institute for Astronomy, University of Hawaii, 2680 Woodlawn Drive, Honolulu, HI 96822, USA}

\author[0009-0002-9727-8326]{X.~Guo}
\affiliation{Department of Astronomy, University of California, Berkeley, CA 94720-3411, USA}

\author[0000-0002-6230-0151]{D.~O.~Jones}
\affiliation{Institute for Astronomy, University of Hawai’i, 640 N. A’ohoku Pl., Hilo, HI 96720, USA}

\author[0000-0002-4571-2306]{J.~Hjorth}
\affil{DARK, Niels Bohr Institute, University of Copenhagen, Jagtvej 128, 2200 Copenhagen, Denmark}

\author[0009-0005-9062-9471]{M.~Kaewmookda}
\affil{National Astronomical Research Institute of Thailand, 260 Moo 4, Donkaew, Maerim, Chiang Mai, 50180, Thailand}

\author[0000-0002-5619-4938]{M.~M.~Kasliwal}
\affiliation{Cahill Center for Astrophysics, California Institute of Technology, MC 249-17, 1216 E California Boulevard, Pasadena, CA, 91125, USA}

\author[0009-0005-1871-7856]{R.~Kaur}
\affiliation{Department of Astronomy and Astrophysics, University of California, Santa Cruz, CA 95064, USA}

\author[0000-0003-2037-4619]{C.~Larison}
\affiliation{Department of Physics and Astronomy, Rutgers, the State University of New Jersey, 136 Frelinghuysen Road, Piscataway, NJ 08854-8019, USA}

\author[0000-0002-2249-0595]{N.~LeBaron}
\affil{Department of Astronomy, University of California, Berkeley, CA 94720-3411, USA}

\author[0000-0003-2736-5977]{H.-Y.~Miao}
\affil{Graduate Institute of Astronomy, National Central University, 300 Zhongda Road, Zhongli, Taoyuan 32001, Taiwan}

\author[0000-0001-6022-0484]{G.~Narayan}
\affiliation{Department of Astronomy, University of Illinois at Urbana-Champaign, 1002 W. Green St., IL 61801, USA}
\affiliation{Center for Astrophysical Surveys, National Center for Supercomputing Applications, Urbana, IL, 61801, USA}

\author[0000-0001-8415-6720]{Y.-C.~Pan}
\affil{Graduate Institute of Astronomy, National Central University, 300 Zhongda Road, Zhongli, Taoyuan 32001, Taiwan}

\author[0000-0001-7488-4337]{S.~H.~Park}
\affil{Department of Physics and Astronomy, Seoul National University, Gwanak-ro 1, Gwanak-gu, Seoul, 08826, South Korea}

\author[0000-0002-1092-6806]{K.~C.~Patra}
\affil{Department of Astronomy and Astrophysics, University of California, Santa Cruz, CA 95064, USA}

\author[0000-0003-3658-6026]{Y.~Qin}
\affiliation{Cahill Center for Astrophysics, California Institute of Technology, MC 249-17, 1216 E California Boulevard, Pasadena, CA, 91125, USA}

\author[0000-0003-4175-4960]{C.~L.~Ransome}
\affil{Steward Observatory, University of Arizona, 933 North Cherry Avenue, Tucson, AZ 85721-0065, USA}
\affil{Center for Astrophysics | Harvard \& Smithsonian, 60 Garden Street, Cambridge, MA 02138-1516, USA}

\author[0000-0002-4410-5387]{A.~Rest}
\affil{Space Telescope Science Institute, Baltimore, MD 21218}
\affiliation{Department of Physics and Astronomy, The Johns Hopkins University, Baltimore, MD 21218, USA}

\author[0000-0003-3643-839X]{J.~Rho}
\affil{SETI Institute}

\author[0000-0003-4725-4481]{S.~Rose}
\affiliation{Cahill Center for Astrophysics, California Institute of Technology, MC 249-17, 1216 E California Boulevard, Pasadena, CA, 91125, USA}

\author[0000-0001-8023-4912]{H.~Sears}
\affiliation{Department of Physics and Astronomy, Rutgers, the State University of New Jersey, 136 Frelinghuysen Road, Piscataway, NJ 08854-8019, USA}

\author[0000-0002-9486-818X]{J.~J.~Swift}
\affiliation{The Thacher School, 5025 Thacher Rd., Ojai, CA 93023, USA}

\author[0000-0002-5748-4558]{K.~Taggart}
\affil{Department of Astronomy and Astrophysics, University of California, Santa Cruz, CA 95064, USA}

\author[0000-0002-5814-4061]{V.~A.~Villar}
\affil{Center for Astrophysics $|$ Harvard \& Smithsonian, Cambridge, MA 02138, USA}
\affil{The NSF AI Institute for Artificial Intelligence and Fundamental Interactions}
\affil{Department of Physics and Kavli Institute for Astrophysics and Space Research, Massachusetts Institute of Technology, 77 Massachusetts Avenue, Cambridge, MA 02139, USA}

\author[0009-0007-7686-5587]{Q.~Wang}
\affiliation{The Thacher School, 5025 Thacher Rd., Ojai, CA 93023, USA}

\author[0000-0002-0632-8897]{Y.~Zenati}
\affil{Space Telescope Science Institute, Baltimore, MD 21218}
\affiliation{Department of Physics and Astronomy, The Johns Hopkins University, Baltimore, MD 21218, USA}
\affiliation{Astrophysics Research Center of the Open University (ARCO), The Open University of Israel, Ra'anana 4353701, Israel}

\author[0000-0002-2093-6960]{H.~Zhou}
\affiliation{The Thacher School, 5025 Thacher Rd., Ojai, CA 93023, USA}

\begin{abstract}

We present multi-wavelength observations of the type II supernova (SN II) 2023ixf during its first two years of evolution. We combine ground-based optical/NIR spectroscopy with {\it Hubble Space Telescope (HST)} far- and near-ultraviolet spectroscopy and {\it James Webb Space Telescope (JWST)} near- and mid-infrared photometry and spectroscopy to create spectral energy distributions of SN~2023ixf at +374 and +620~days post-explosion, covering a wavelength range of $\sim 0.1-30~\mu$m. The multi-band light curve of SN~2023ixf follows a standard radioactive decay decline rate after the plateau until $\sim 500$~days, at which point shock powered emission from ongoing interaction between the SN ejecta and circumstellar material (CSM) begins to dominate. This evolution is temporally consistent with 0.3-10~keV X-ray detections of SN~2023ixf and broad ``boxy'' spectral line emission, which we interpret to signal reprocessing of shock luminosity in a cold dense shell located between forward and reverse shocks. Using the expected absorbed radioactive decay power and the detected X-ray luminosity, we quantify the total shock powered emission at the +374 and +620~day epochs and find that it can be explained by nearly complete thermalization of the reverse shock luminosity as SN~2023ixf interacts with a continuous, ``wind-like'' CSM with a progenitor mass-loss rate of $\dot M \approx 10^{-4}$~\mdot\ ($v_w = 20 \pm 5~\kms$). Additionally, we construct multi-epoch spectral models from the non-LTE radiative transfer code {\tt CMFGEN}, which contain radioactive decay and shock powers, as well as dust absorption, scattering, and emission. We find that models with shock powers of $L_{\rm sh} = (0.5-1) \times 10^{40}$~erg~s$^{-1}$ and $\sim(0.5 - 1) \times 10^{-3}~\Msun$ of silicate dust in the cold dense shell and/or inner SN ejecta can effectively reproduce the global properties of the late-time ($>300$~days) UV-to-IR spectra of SN~2023ixf.  

\end{abstract}

\keywords{Type II supernovae (1731) --- Red supergiant stars (1375) --- Circumstellar matter (241) --- Ultraviolet astronomy (1736) --- Spectroscopy (1558) --- Shocks (2086) -- X-ray astronomy (1810) }

\section{Introduction} \label{sec:intro}

Supernova (SN) 2023ixf is a type II supernova (SN II) that occurred on 2023-05-18 in Messier 101 at a distance of $6.85\pm 0.15$~Mpc \citep{riess22}. SN~2023ixf is one of the closest SN in the last decade and, consequently, was reported to the Transient Name Server (TNS) within $\sim$1 day of first light by Koichi Itagaki \citep{Itagaki23} on 2023 May 19 17:27:15 (MJD 60083.73). The earliest reported detection of SN~2023ixf was on MJD 60082.85 at $17.1\pm0.1$~mag (-12.1~mag) in $r$-band following a deep upper limit of $>20.4$~mag ($< -8.78$~mag) on MJD 60082.66 \citep{Mao23, Li24}. Consequently, an estimate of the time of first light is MJD $60082.76 \pm 0.01$, which is adopted throughout this work. 

The early-time spectra of SN2023ixf exhibit emission lines from \ion{H}{i}, \ion{He}{i/ii}, \ion{C}{iv}, and \ion{N}{iii/iv/v}, superimposed on a hot, blue spectral continuum \citep{wjg23, Bostroem23, Teja23, Zhang23}. These lines, commonly referred to as ``flash'' or ``IIn-like'' features, likely originate from the sustained photoionization of dense circumstellar material (CSM) located ahead of the forward shock \citep{galyam14, Khazov16, yaron17, dessart17, terreran22, bruch21, bruch23, Dessart23, wjg24}. Much like Type IIn supernovae, the emission line profiles consist of a narrow core accompanied by broad Lorentzian ``wings'' that extend up to $\sim$1000~$\kms$, a signature of electron scattering within the ionized, optically thick CSM \citep{Chugai01, Dessart09, dessart17, Huang18}. Additionally, SN~2023ixf has the earliest and highest cadence UV spectroscopy \citep{Teja23, Zimmerman24}, spectropolarimetry \citep{Vasylyev23, Singh24, Shrestha25, Vasylyev25} and high resolution spectroscopy \citep{smith23, Zimmerman24, Dickinson25} of any CSM-interacting SN~II to date. 

Multi-band photometry of SN~2023ixf began within hours of first light, capturing shock emergence from a dense, compact CSM at $<10^{14}$~cm. The early light curve shows a two-component rise: the first $\sim$12 hours require a distinct power-law fit, while the later evolution ($\delta t > 12$h) follows a typical $F_{\nu} \propto t^2$ trend \citep{Hosseinzadeh23}. During its first day, SN~2023ixf exhibited a rapid red-to-blue color evolution, common among CSM-interacting SNe II with IIn-like features (e.g., SN~2024ggi; \citealt{wjg24b, Shrestha24}), likely marking shock breakout from the innermost CSM. This coincided with a nearly constant blackbody radius and rising temperature \citep{Zheng25, Zimmerman24}. \cite{Li24} attributes the early red colors to dust destruction during breakout, consistent with the IR excess detected by NEOWISE-R at $\delta t = 3.6$~days \citep{VanDyk24}.

The dust-enshrouded red supergiant (RSG) progenitor star of SN~2023ixf was clearly detected in multiple {\it HST} optical bands, in Channels 1 and 2 of {\it Spitzer}, and in $J$ and $K$ bands from Gemini-North and MMT. Estimates of the progenitor’s Zero Age Main Sequence (ZAMS) mass vary widely, ranging from $\sim$8–20~$\Msun$ \citep{Kilpatrick23, Jencson23, Niu23, Soraisam23, VanDyk23, Qin23, Pledger23, Xiang24, Ransome24} and no evidence of a binary companion was found, though only secondary stars with masses above $6.4~\Msun$ could be excluded. Additional progenitor mass constraints were made from model matching to the SN~2023ixf light curve \citep{Moriya24,Singh24, Bersten24, Hsu24, Forde25, Cosentino25} and nebular spectroscopy \citep{Ferrari24, Fang25, Kumar25, Michel25, Li25,Folatelli25, wjg25}. Notably, {\it Spitzer} data showed long-term variability with a pulsation period of $\sim$1000 days \citep{Kilpatrick23, Jencson23, Soraisam23}. Given the early-time CSM interaction observed in SN~2023ixf, several studies searched for pre-explosion outbursts, common in some CSM-interacting SNe~II (e.g., \citealt{Strotjohann21, wjg22}), but found no precursor activity in archival X-ray/UV/optical/NIR data \citep{Panjkov24, Neustadt24, Ransome24, Dong23, Rest25, Flinner23}.

\begin{figure*}
\centering
\includegraphics[width=\textwidth]{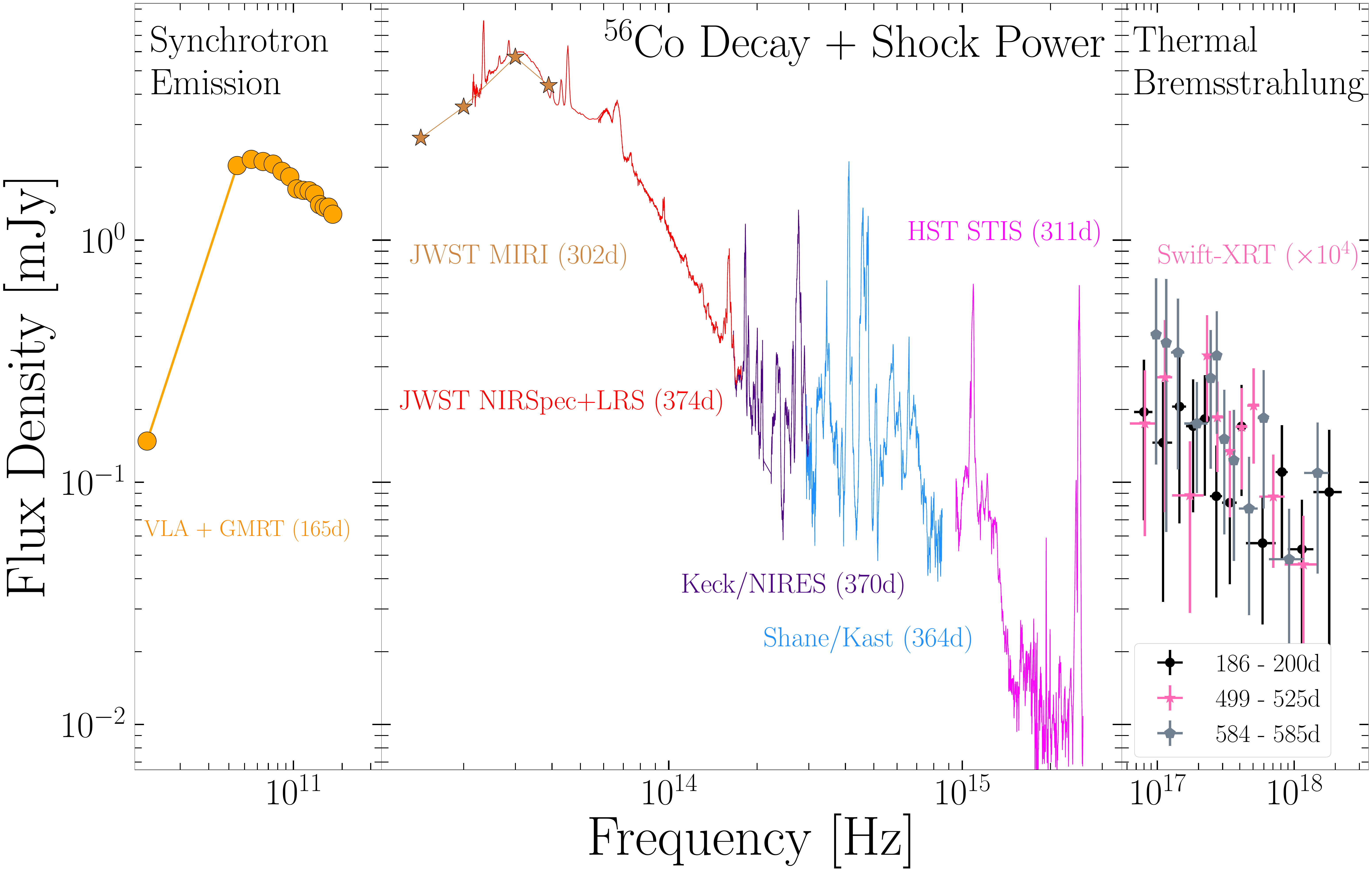}
\caption{ Late-time spectral energy distribution of SN~2023ixf covering radio, infrared, optical, UV and X-ray wavelengths at $\delta t \approx 165 - 585$~days. Emission mechanisms and power sources are labeled at the top of the figure. 
    \label{fig:SED_all} }
\end{figure*}

X-ray observations of SN2023ixf began with {\it Swift}-XRT ($\delta t \approx 1$~day), followed by {\it NuSTAR} (3–79~keV; $\delta t = 4.4–58.4$d), {\it Chandra} (0.5–8~keV; $\delta t = 11.5–86.7$~days), and {\it XMM-Newton} (0.3–10~keV; $\delta t = 9.0–58.2$~days). This unprecedented soft and hard X-ray coverage enabled the first precise temperature constraints for a SN~II with IIn-like features \citep{Grefenstette23, Chandra24, Panjkov24, Nayana25}. The X-ray emission, dominated by the forward shock, is best fit by an absorbed bremsstrahlung model \citep{Nayana25, Chandra24, Panjkov24}, peaking at $\sim$$10^{40}$~erg~s$^{-1}$ within a week. Detection of Fe K$\alpha$ and high intrinsic $N_\mathrm{H}$ confirms early X-ray absorption by dense, confined CSM \citep{Grefenstette23, Nayana25} and the observed rising in X-ray flux reflects decreasing photoelectric absorption as CSM density also decreases and the ionization fraction potentially rises. A high-cadence radio campaign began at $\delta t = 2.6$~d with SMA at mm wavelengths, though no emission was detected until $\delta t = 29.2$~d with the VLA at 10~GHz \citep{Berger23, Matthews23}. Additional broad-band radio spectral coverage was provided by GMRT, LOFAR, NOEMA, the Japanese and Korean VLBI Networks, and EVN \citep{Nayana25, Timmerman24, Iwata25, Lee24}. The radio emission, produced by synchrotron radiation from the forward shock, is initially mostly suppressed by free–free absorption, which weakens as the CSM density decreases at larger shock radii \citep{Nayana25}. Despite potentially contradictory mass-loss rate estimates between optical spectroscopy and X-ray observations within the first week post-explosion, long-term X-ray and radio monitoring provides a consistent mass-loss rate of $\sim 2\times 10^{-4}$~\mdot\ at shock radii $>10^{15}$~cm \citep{Nayana25, Panjkov24}. A complete review of all observations and analysis of SN~2023ixf to date is presented in \cite{wjg23ixf}.

In this paper, we present multi-wavelength observations and analysis of the continuous 0.1-30~$\mu$m SN~2023ixf SED at late-time phases (Fig. \ref{fig:SED_all}). In \S\ref{sec:obs}, we present observations and reduction techniques. In \S\ref{sec:analysis}, we present an analysis of the radioactive decay and shock powered emission in the SN~2023ixf SED. Conclusions are then presented in \S\ref{sec:conclusion}. All phases reported in this paper are with respect to the adopted time of first light ($60082.757 \pm 0.097$) and are in rest-frame days, as denoted by $\delta t$. We adopt a redshift of $z = 0.000804$ \citep{perley23}, and Milky Way and host galaxy reddening values of $E(B-V)_{\rm MW} = 0.008$~mag and $E(B-V)_{\rm host} = 0.033 \pm 0.010$~mag \citep{wjg23}.

\section{Observations} \label{sec:obs}

\subsection{Photometric Observations}\label{SubSec:Phot}

\sn{} was observed with the Pan-STARRS telescope \citep[PS1/2;][]{Kaiser2002, Chambers2017} between 2023-05-21 and 2025-06-02 in $grizy$-bands through the Young Supernova Experiment (YSE) \citep{Jones2021}. Data storage/visualization and follow-up coordination was done through the YSE-PZ web broker \citep{Coulter22, Coulter23}. The YSE photometric pipeline is based on {\tt photpipe} \citep{Rest+05}, which relies on calibrations from \citep{Magnier20a, waters20}. Each image template was taken from stacked PS1 exposures, with most of the input data from the PS1 3$\pi$ survey. All images and templates were resampled and astrometrically aligned to match a skycell in the PS1 sky tessellation. An image zero-point is determined by comparing PSF photometry of the stars to updated stellar catalogs of PS1 observations \citep{flewelling16}. The PS1 templates are convolved with a three-Gaussian kernel to match the PSF of the nightly images, and the convolved templates are subtracted from the nightly images with {\tt HOTPANTS} \citep{becker15}. Finally, a flux-weighted centroid is found for the position of the SN in each image and PSF photometry is performed using ``forced photometry": the centroid of the PSF is forced to be at the SN position. The nightly zero-point is applied to the photometry to determine the brightness of the SN for that epoch.

We obtained $ugri$ imaging of SN\,2023ixf with the Las Cumbres Observatory (LCO) 1\,m telescopes from 2023-05-20 to 2024-09-24 (Programs NSF2023A-011, NSF2023A-015, NSF2023B-004, ANU2024A-004, ANU2024B-003, and ANU2025A-003; PIs Foley, Kilpatrick, Auchettl).  After downloading the {\tt BANZAI}-reduced images from the LCO data archive \citep{mccully18}, we used {\tt photpipe} \citep{Rest+05} to perform {\tt DoPhot} PSF photometry \citep{Schechter+93}. All photometry was calibrated using PS1 stellar catalogs described above with additional transformations to SDSS $u$-band derived from \citet{finkbeiner16}. For additional details on our reductions, see \citet{kilpatrick18}. We also observed SN\,2023ixf with the Lulin 1\,m and Thacher \citep{Swift22} telescopes in $griz$ bands from 2023-05-21 to 2025-04-20.  Standard calibrations for bias and flat-fielding were performed on the images using {\tt IRAF}, and we reduced the calibrated frames in {\tt photpipe} using the same methods described above for the LCO images. We obtained NIR photometry using the Wide-field InfraRed Camera (WIRC; \citealt{Wilson03}) on the 200-inch telescope at Palomar Observatory. The WIRC data were reduced using standard methods for dark subtraction, flat-fielding, sky subtraction, astrometric and photometric calibration \citep{De2021}.

The Ultraviolet Optical Telescope (UVOT; \citealt{Roming05}) onboard the Neil Gehrels \emph{Swift} Observatory \citep{Gehrels04} observed \sn{} from 2023-05-19 to 2025-02-23 ($\delta t = 1.43 - 646.3$~days). We supplement the published early-time UVOT photometry from \cite{Zimmerman24} with reductions at $\delta t > 50$~days. We performed aperture photometry with a 5$\arcsec$ region radius with \texttt{uvotsource} within HEAsoft v6.33 \citep{HEAsoft}\footnote{We used the most recent calibration database (CALDB) version.}, following the standard guidelines from \cite{Brown14}\footnote{\url{https://github.com/gterreran/Swift_host_subtraction}}. In order to remove contamination from the host galaxy, we employed pre-explosion images to subtract the measured count rate at the location of the SN from the count rates in the SN images and corrected for point-spread-function (PSF) losses following the prescriptions of \cite{Brown14}. 

\begin{figure*}
\centering
\includegraphics[width=\textwidth]{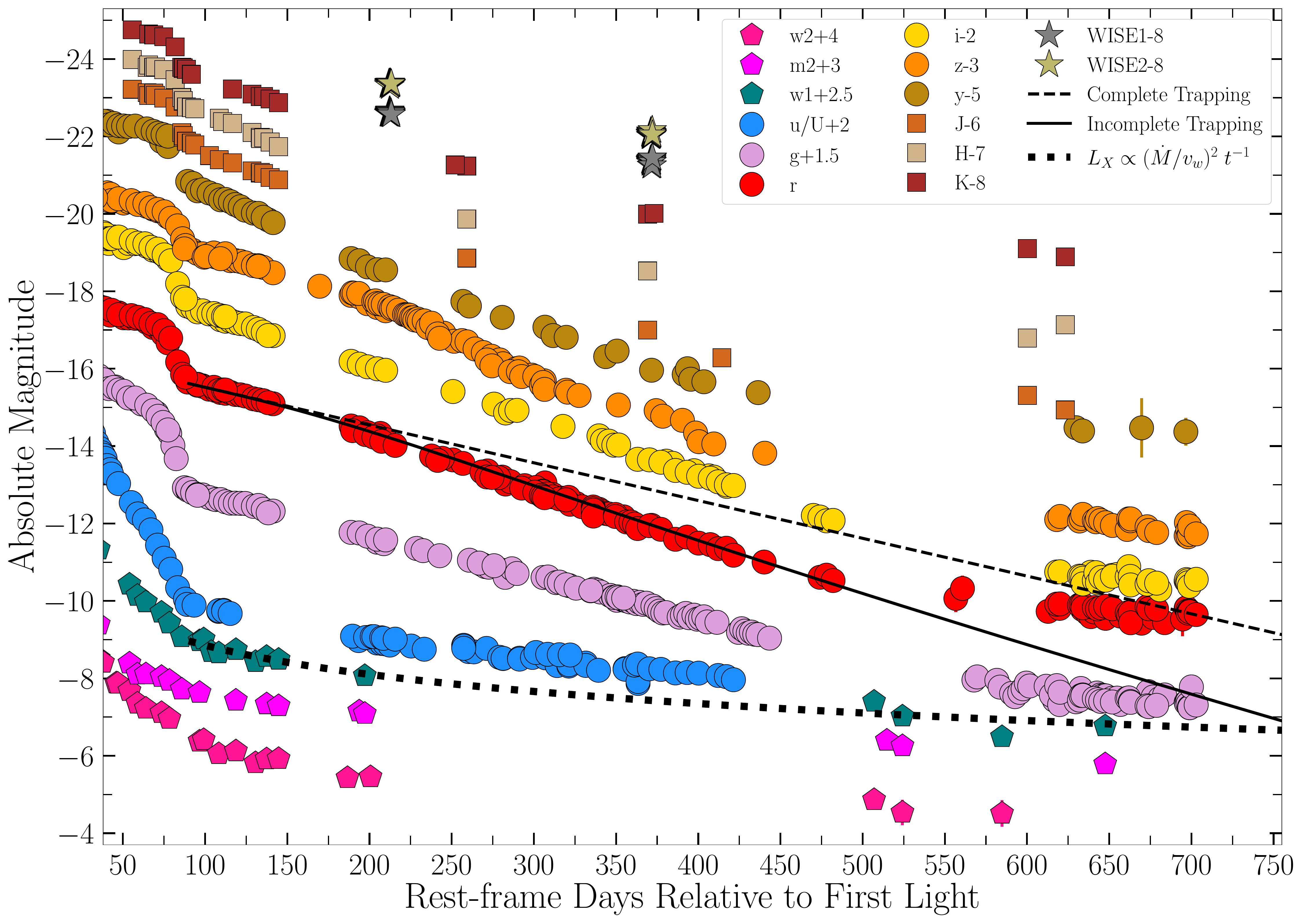}
\caption{ Multi-band light curve of SN~2023ixf extending to $\delta t \approx 700$~days and covering UV (polygons), optical (circles), NIR (squares) and MIR (stars) wavelengths. Radioactive decay power model decline rates shown as dashed (complete $\gamma$-ray trapping) and solid (incomplete $\gamma$-ray trapping) black lines. Decline rate for free-free X-ray emission from CSM-interaction shown as dotted line. The dominance of shock powered emission from CSM-interaction over absorbed radioactive decay power is observed as excess emission in UV filters at $> 200$~days and optical/NIR filters at  $> 500$~days. 
\label{fig:LC_all} }
\end{figure*}

In addition to our observations, we include $gri$-band photometry from the Zwicky Transient Facility (ZTF; \citealt{bellm19, graham19}) forced-photometry service \citep{Masci19}. We also include multi-band photometry presented in \cite{wjg23, VanDyk24, Hsu24, Zimmerman24, Singh24} in our analysis. Furthermore, we include {\it JWST} MIRI imaging of SN~2023ixf from 2024-03-15 (JWST-GO-3921, PI: Fox) and 2025-01-27 (JWST-GO-5290, PI: Ashall), which includes observations in F770W, F1000W, F1500W, F1800W, F2100W, and F2500W filters. We performed PSF photometry on all filters and epochs using the {\tt spacephot} photometry package and the PSF models from {\tt WebbPSF}.\footnote{\url{https://space-phot.readthedocs.io/en/latest/}} All photometric measurements are consistent with those derived in \cite{Medler25}. The complete UVOIR light curve of SN~2023ixf is shown in Figure \ref{fig:LC_all}. 

% \begin{figure*}
% \centering
% \includegraphics[width=0.99\textwidth]{SED_radio_to_xray.pdf}
% \caption{
% \label{fig:SED_all} }
% \end{figure*}

\begin{figure*}
\centering
\includegraphics[width=\textwidth]{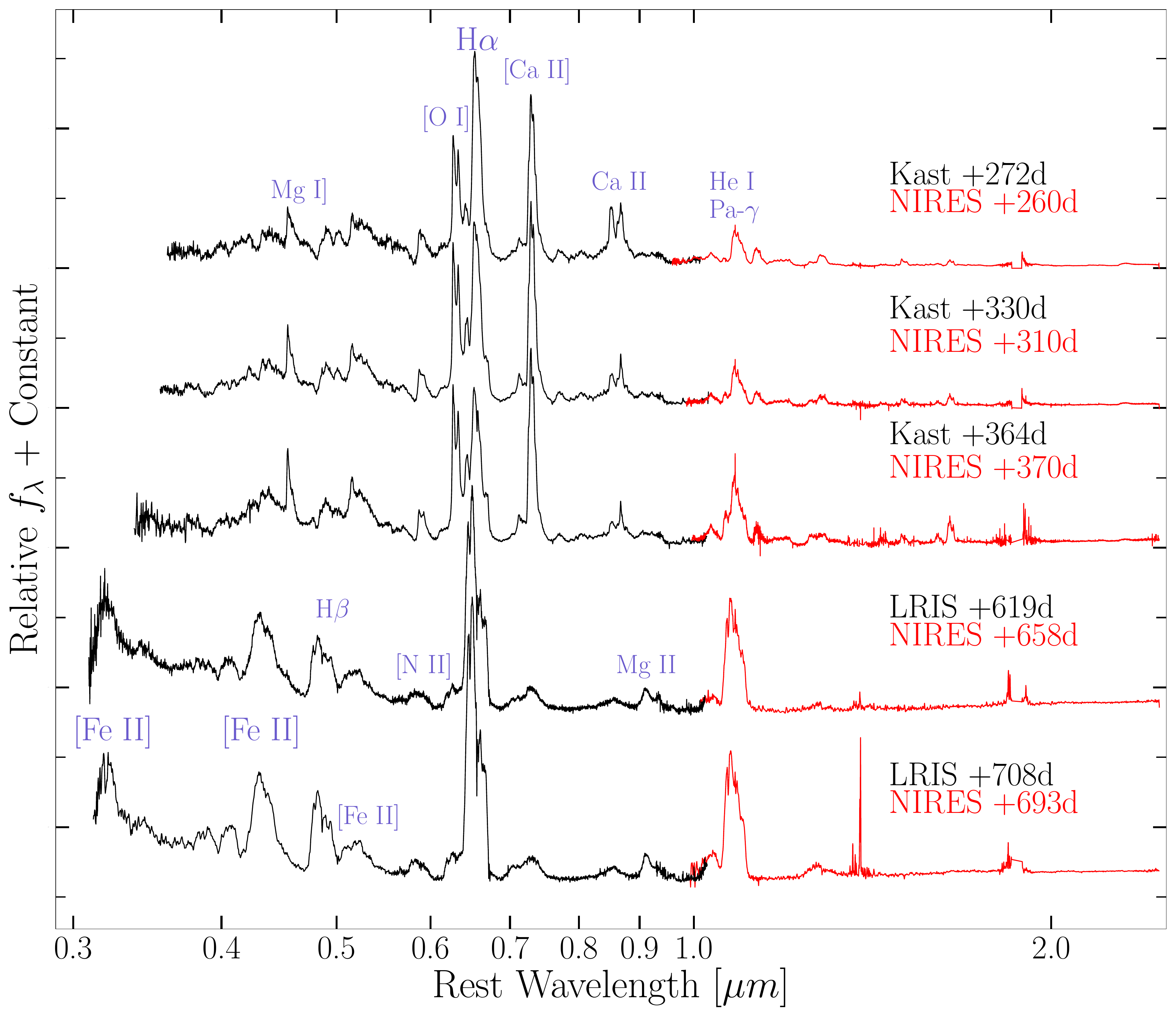}
\caption{ Time-series optical/NIR (black/red) spectra of SN~2023ixf spanning $\delta t = 272 - 708$~days. Ion identifications shown in blue are based on simulations from \cite{Dessart23b}. Spectra at $\delta t > 619$~days are dominated by broad, ``boxy'' line profiles derived from emission in the dense shell and inner ejecta as a radiative reverse shock is powered by ongoing CSM-interaction.  
\label{fig:spec_all} }
\end{figure*}

\begin{figure*}
\centering
\subfigure{\includegraphics[width=0.52\textwidth]{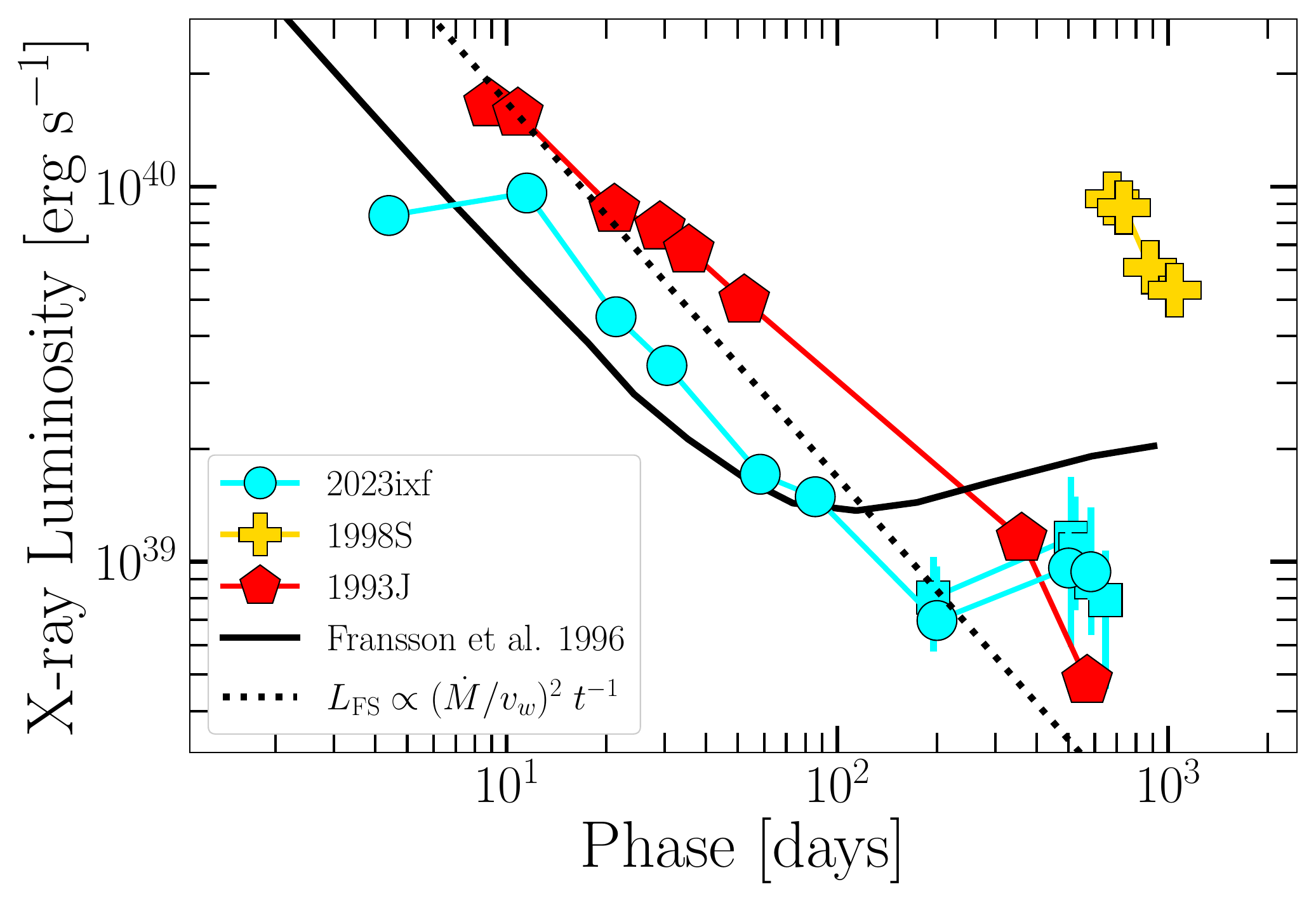}}
\subfigure{\includegraphics[width=0.47\textwidth]{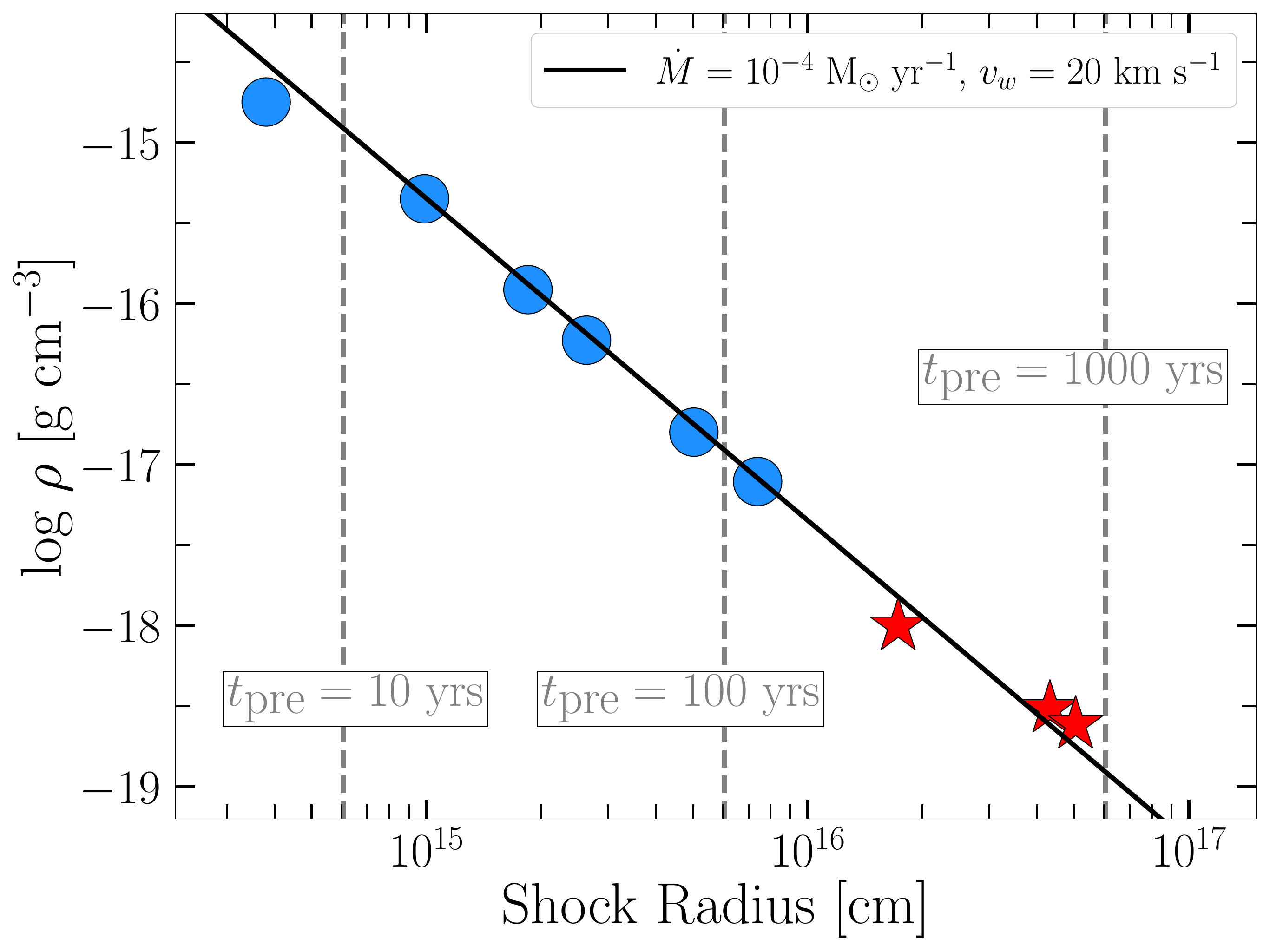}}
\caption{{\it Left:} Unabsorbed 0.3-10~keV light curve of SN~2023ixf (cyan circles from method 1 and cyan squares from method 2; \S\ref{SubSec:Xray}) compared to SNe~1993J (red polygons) and 1998S (yellow plus signs). Black solid line is the multi-shock model for SN~1993J from \cite{Fransson96} where rising flux at $>100$~days is from the emerging RS. Black dotted and dashed lines represent the analytic predictions for free-free emission (e.g., Eqn. \ref{eqn:Lff}) for the forward shock luminosity assuming $\dot M = 10^{-4}$~\mdot\ ($v_w = 20~\kms$). {\it Right:} CSM density measurements from \cite{Nayana25} (blue circles) and modeling of late-time X-ray spectra (red stars). The derived CSM density profile continues to trace a steady-state mass loss rate of $\dot M = 10^{-4}$~\mdot\ (solid black line). Dashed grey lines represent look-back times assuming a shock velocity of $10^{4}~\kms$ and CSM velocity of $20~\kms$.  \label{fig:Xray} }
\end{figure*}

% \begin{figure}[t!]
% \centering
% \includegraphics[width=0.49\textwidth]{SED_dust_fits.pdf}
% \caption{\label{fig:dust} }
% \end{figure} The RS model has been self-consistently corrected for photo-electric absorption in the CDS and the displayed light curve is for the predicted luminosity at 1~keV.

% \begin{figure*}
% \centering
% \subfigure{\includegraphics[width=0.49\textwidth]{NeII_CMFGEN_ni56.pdf}}
% \subfigure{\includegraphics[width=0.49\textwidth]{NeII_CMFGEN_bol.pdf}}
% \caption{\label{fig:LC_bol} }
% \end{figure*}

\subsection{Spectroscopic Observations}\label{SubSec:Spec}

We obtained late-time spectra of SN~2023ixf with the Kast spectrograph on the 3-m Shane telescope at Lick Observatory \citep{KAST}. For all of these spectroscopic observations, standard CCD processing and spectrum extraction were accomplished with \textsc{IRAF}\footnote{\url{https://github.com/msiebert1/UCSC\_spectral\_pipeline}}. The data were extracted using the optimal algorithm of \citet{1986PASP...98..609H}.  Low-order polynomial fits to calibration-lamp spectra were used to establish the wavelength scale and small adjustments derived from night-sky lines in the object frames were applied. Additional optical spectra were obtained with Keck/LRIS \citep{oke95} and reduced with {\tt Lpipe} \citep{Perley19}. Near-infrared (NIR) spectra were obtained with the $R\approx2700$ Near-Infrared Echelle Spectrograph (NIRES; \citealt{Wilson04}) located on Keck-II. The data were reduced using a custom version of the IDL based reduction package {\tt Spextool} \citep{Cushing04} modified for use with NIRES as well as the {\tt Pypeit} spectral reduction pipeline \citep{Prochaska20}. For the {\tt Spextool} reduction, we used {\tt xtellcor} \citep{vacca03} to correct for telluric features in our spectrum using an A0 standard star observed close in airmass and time to our target. NIRES data from $\delta t = 259.8 - 658.8$~days were obtained through the Keck Infrared Transient Survey (KITS; \citealt{KITS}).

We observed SN~2023ixf on 2024-12-31 starting at 15:44 UT with both the Medium and Small slicers of the Keck Cosmic Web Imager (KCWI; blue channel, \citealt{Morrissey18}) and the Keck Cosmic Reionization Mapper (KCRM; red channel) mounted on the Keck II telescope. For the Medium slicer observations, the blue channel was configured with the BL grating centered at 4500~\AA\ and the red channel used the RL centered at 7150~\AA. We acquired one 850s image in the blue channel and 2x300s in the red channel, both channels configured with the 2x2 binning. For the Small slicer, we used the BL grating centered at 4500~\AA\ and a 2x2 binning to acquire 2x650s images, while for the red channel we used the RH1 grating ($R \approx 13000$) centered at 6520~\AA\ and a 1x1 binning to target the H$\alpha$ line of SN~2023ixf. 

Far- and near-UV spectra of SN~2023ixf were obtained on 2024-03-24 ($\delta t = 311$~days) and 2025-01-26 ($\delta t = 619$~days) with the {\it Hubble Space Telescope (HST)} using the Multi-Anode MicroChannel Array (MAMA) detectors on the Space Telescope Imaging Spectrograph (STIS) through HST programs 17497 (PI Valenti) and 17772 (PI Bostroem) \cite{Bostroem24, Bostroem25}. Standard bias subtraction, flat-fielding and cosmic ray rejection were performed automatically before downloading from the Mikulski Archive for Space Telescopes (MAST). Wavelength calibration, flux calibration and 1-D extraction was performed using the {\tt calstis} pipeline routine. Near- and mid-infrared spectroscopy of SN~2023ixf were obtained on 2024-01-26, 2024-05-26, and 2025-01-08 ($\delta t = 252.8 - 600.6$~days) (JWST-DD-4575, JWST-GO-5290; PI: Ashall) with the {\it James Webb Space Telescope (JWST)} using NIRSpec with G235M-F170LP and G395M-F290LP grating/filter combinations \citep{Jakobsen22} and MIRI/LRS \citep{Kendrew15}. These data were reduced using the {\tt jwst}\footnote{\url{https://github.com/spacetelescope/jwst}} pipeline, which performs standard bias subtraction, flat-fielding, wavelength and flux calibrations, and spectral extraction. The complete optical/NIR spectral sequence is shown in Figure \ref{fig:spec_all} and a log of all spectral observations utilized in this paper is provided in Table \ref{tab:spec_all}. All photometric and spectroscopic data will be made public online.\footnote{\url{https://github.com/wynnjacobson-galan/SN2023ixf}}

\subsection{X-ray Observations with \emph{Swift}-XRT}\label{SubSec:Xray}

The X-Ray Telescope (XRT, \citealt{burrows05}) on board the \emph{Swift} spacecraft \citep{Gehrels04} observed the field of SN~2023ixf from 2023-05-19 to 2025-02-23 ($\delta t = 1.43 - 646.3$~days). Observations across multiple X-ray telescopes, including XRT during the first $\sim$200 days of SN~2023ixf have been analyzed and published in recent studies (e.g., \citealt{Grefenstette23, Panjkov24, Chandra24, Zimmerman24, Nayana25}). We focus on the XRT epochs from $\delta t = 200 - 646$~days, which cover the range of late-time UVOIR observations discussed in \S\ref{SubSec:Phot} \& \S\ref{SubSec:Spec}. We analyzed the data using HEAsoft v6.33 and followed the prescriptions detailed in \cite{margutti13}, applying standard filtering and screening. Below we outline our first method for modeling and analyzing the XRT data with {\tt Xspec}, but for completeness we also outline a secondary modeling method in Appendix \S\ref{sec:xrayextra}. For this first method, we chose to merge the event files and use the combined epoch for analysis of the X-ray spectrum in the following date ranges: 2023-11-21 to 2023-12-05 ($t_{\rm exp} = 16.9$~ks), 2024-09-29 to 2024-10–24 ($t_{\rm exp} = 13.8$~ks), and 2024-12-23 to 2024-12-24 ($t_{\rm exp} = 9.1$~ks). In all merged images, a bright source of X-ray emission is clearly detected at the SN location with significance of $>3\sigma$ against the background.

\begin{figure*}
\centering
\subfigure{\includegraphics[width=0.54\textwidth]{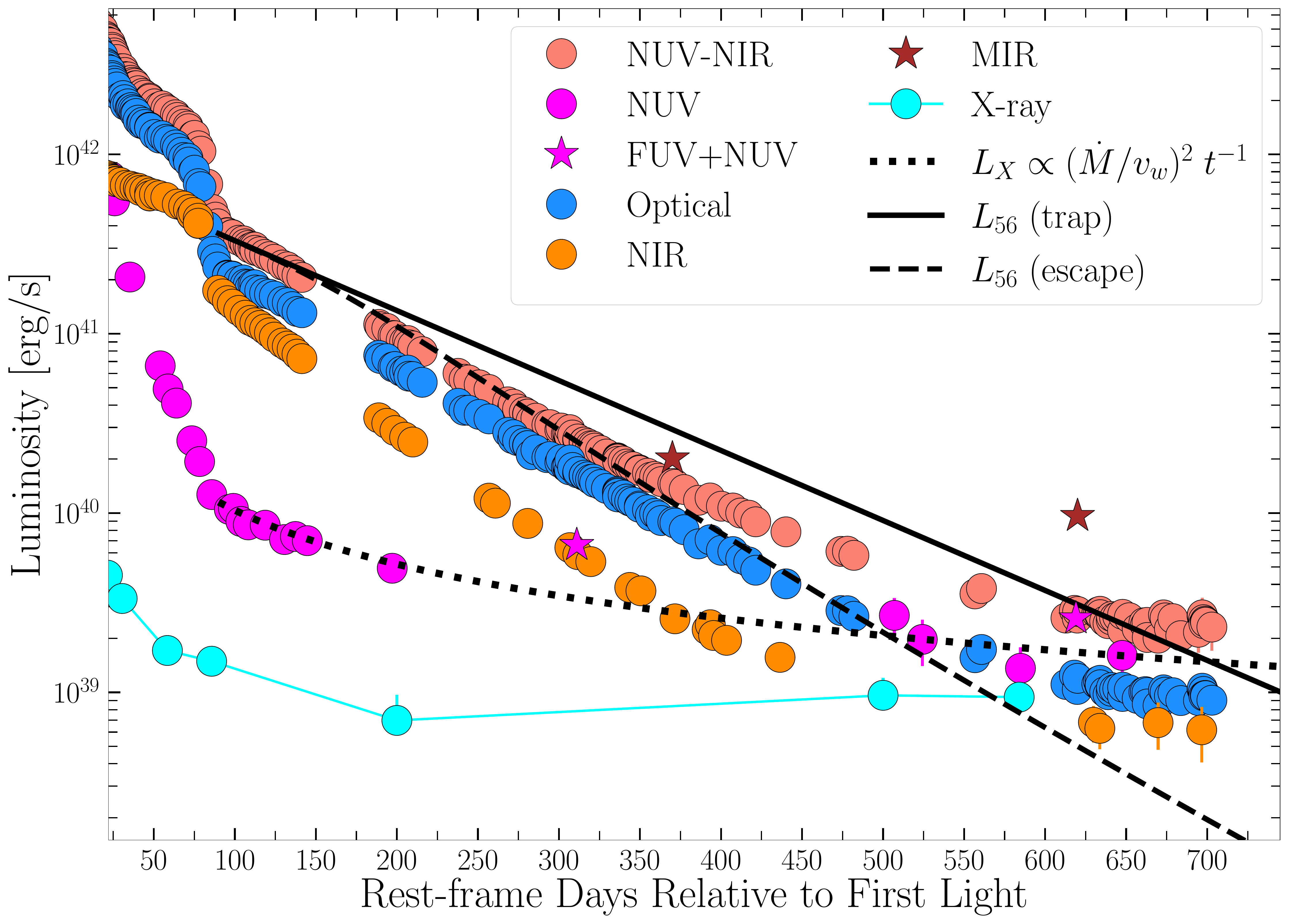}}
\subfigure{\includegraphics[width=0.45\textwidth]{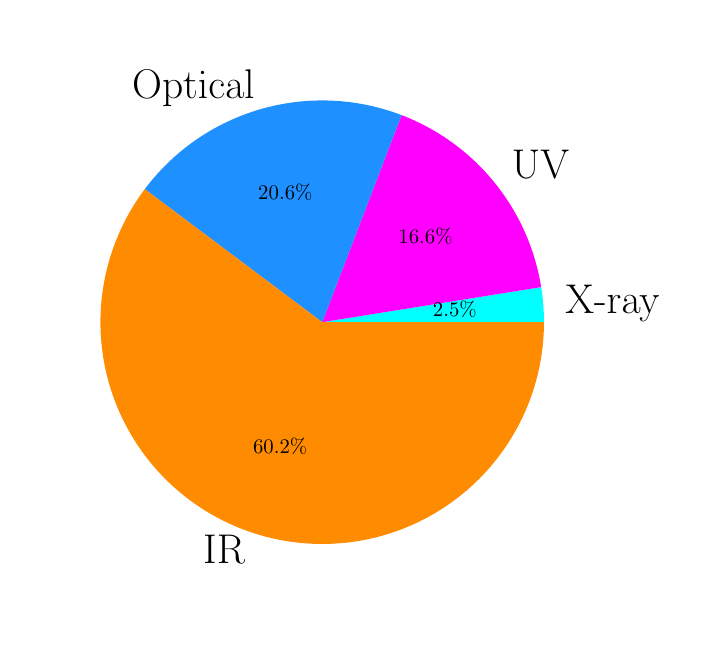}}
\caption{ {\it Left:} Pseudobolometric light curves of SN~2023ixf constructed using NUV-NIR bands ($0.16 - 2.5~\mu$m, salmon circles), NUV ($0.16 - 0.3~\mu$m, magenta circles), FUV-NUV ($0.1 - 0.3~\mu$m, magenta stars), optical ($0.3 - 1.0~\mu$m, blue circles), NIR ($1.0 - 2.5~\mu$m, orange circles), and MIR ($2.5 - 30.0~\mu$m, orange circles). Unabsorbed 0.3-10~keV X-ray light curve shown as cyan circles. {\it Right:} Fractional flux emitted in UV ($0.1 - 0.3~\mu$m, magenta), optical ($0.3 - 1.0~\mu$m, blue), IR ($1 - 30~\mu$m, orange), and X-ray ($0.3 - 10$~keV, cyan) wavelengths at $\delta t = 374$~days. \label{fig:LC_bol} }
\end{figure*}

From each merged event file, we extracted a spectrum using a 25$\arcsec$ region centered at the location of SN~2023ixf and correct for background emission with a 100$\arcsec$ source-free region. Similar to \cite{Grefenstette23}, \cite{Chandra24}, \cite{Panjkov24}, and \cite{Nayana25}, we use {\tt Xspec} to model each 0.3-10~keV spectra with an absorbed thermal bremsstrahlung model ({\tt tbabs*ztbabs*bremss}), which includes solar abundances \citep{Asplund09}, and a line-of-sight hydrogen column density of $N_{\rm H,MW} = 7.9 \times 10^{20}$~cm$^{-2}$. Furthermore, each model spectrum includes an additional power law component in order to account for the contaminating source presented in \cite{Nayana25}. We fit each X-ray spectrum with two separate models: (1) two temperature components for the forward shock (FS) and reverse shock (RS) and (2) single temperature model for only the FS. We fix the temperatures in each spectral fit so that the FS and RS temperature evolves as $T_{\rm FS} \propto t^{-0.5}$ \citep{Nayana25} and $T_{\rm RS}  = T_{\rm FS} / (n-3)^2$, assuming an ejecta density profile power-law index of $n = 12$ \citep{Chevalier17, Chandra24}. As shown in Appendix Figure \ref{fig:XRT}, the relatively small number of counts and the lower \emph{Swift}-XRT sensitivity at $<$1~keV inhibits a meaningful constraint on intrinsic $N_H$ ahead of the FS, which is likely $<10^{21}$~cm$^{-2}$ based on the measured decline rate in \cite{Nayana25}. When fitting with a RS model component, we invoke a time-dependent CDS column density of $N_{\rm CDS} \approx 10^{25} / (t/d)$~cm$^{-2}$ assuming a ``wind-like'' CSM density profile (e.g., see \citealt{Chevalier17, Nayana25}), which is fixed in each model fit. To determine whether an additional temperature component for the RS is necessary, we calculate the Bayesian Information Criterion (BIC) for each model fit to the SN~2023ixf X-ray spectrum. As shown in Table \ref{tab:XRT}, there is no statistical evidence that a RS emission component is needed to best fit the X-ray data. Furthermore, we report the 3$\sigma$ uncertainties on the normalization parameter for the RS component and find that the best-fit value is consistent with zero for the first two epochs and unphysical for the third epoch. All X-ray modeling parameters from this method and that described in \S\ref{sec:xrayextra} are reported in Table \ref{tab:XRT}.

In Figure \ref{fig:Xray} we present the unabsorbed 0.3 - 10~keV X-ray light curve from \cite{Nayana25} and the additional luminosities derived from our analysis using a single temperature model fit. While the $\delta t \approx 200$~day data are consistent with the $t^{-1}$ decline rate expected from pure FS emission, the flattening of the $0.3 - 10$~keV X-ray light curve at $\delta t > 500$~days may be indicative of emerging RS emission and/or increasing CSM density. To quantify the CSM density at shock radii $>10^{16}$~cm, we use the normalization of the X-ray spectrum to calculate the total emission measure (EM) at each epoch, which can then be converted to unshocked CSM density (e.g., see Eqn. 2 in \citealt{Brethauer22}). We adopt all the same parameters as \cite{Nayana25} when applying this EM formalism. As shown in Figure \ref{fig:Xray}, the CSM density continues to decline as $\rho \propto r^{-2}$, which can be matched by a steady-state mass loss rate of $\dot M = 2\times 10^{-4}$~\mdot\ ($v_w = 25~\kms$). Overall, we find no evidence for a RS contribution to the observed $0.3 -10$~keV X-ray luminosity shown in Figure \ref{fig:Xray}.

\section{Analysis}\label{sec:analysis}

\subsection{Radioactive Decay Power}\label{subsec:Lrad}

%After a $\sim$83~day light curve photospheric phase (see Figure \ref{fig:LC_all}), the photometric evolution of SN~2023ixf became dominated by the radioactive decay from the standard ${}^{56}\rm{Ni} \rightarrow {}^{56}\rm{Co} \rightarrow {}^{56}$Fe chain \citep{arnett82}.

As shown in Figure \ref{fig:LC_all}, the photospheric phase persists as a linearly-declining plateau in the SN~2023ixf light curve for $\sim$83~days. Following the fall from the plateau, the photometric evolution of SN~2023ixf becomes consistent with radioactive decay from the standard ${}^{56}\rm{Ni} \rightarrow {}^{56}\rm{Co} \rightarrow {}^{56}$Fe chain \citep{arnett82}. Modeling of the nebular decline rate has suggested $M(^{56} \rm Ni) \approx 0.05-0.07~\Msun$ and incomplete $\gamma$-ray trapping with a characteristic timescale of $t_{\gamma} \approx 250- 300$~days \citep{Zimmerman24, Singh24, Li25}. To confirm these parameters, we construct two variations of the post-plateau bolometric light curve, which are then fit with analytic formalisms for radioactive decay power. First, we construct a pseudobolometric light curve using $UBVgri$ filters and fit this light curve with the pseudobolometric light curve of SN~1987A derived with the same filter combination (e.g., see Eqn. 1 of \citealt{wjg25}), which yields $M(^{56} \rm Ni) = 0.059 \pm 0.001~\Msun$ and $t_{\gamma} = 268.7 \pm 3.6$~days. We note that consistent values are found when this method is applied to the more confined $gri$ filter combination. Second, we construct a pseudobolometric light curve from the combination of $UBgVrizyJHK$ filters and which covers $0.3-2.5~\mu$m. Given the emergence of shock powered emission at late-times (\S\ref{subsec:Lsh}), we model this pseudobolometric light curve at $\delta t < 400$~days and find $M(^{56} \rm Ni) = 0.056 \pm 0.00040~\Msun$ and $t_{\gamma} = 260.6 \pm 3.5$~days. We chose to exclude UV and mid-IR photometry in this light curve given that (1) $\sim 95~\%$ of SN~II luminosity is expected to be emitted between $0.3-2.5~\mu$m at $\delta t = 200 - 400$~days \citep{Dessart25IR} if it arises from decay power only and (2) the UV excess from persistent shock powered emission shown in Figure \ref{fig:LC_bol} would lead to an overestimate of $M(^{56} \rm Ni)$. Based on the average of both modeling attempts, we choose to adopt $M(^{56} \rm Ni) = 0.0580 \pm 0.0010~\Msun$ and $t_{\gamma} = 264.6 \pm 2.5$~days when calculating the contribution of absorbed radioactive decay power to the observed luminosity of SN~2023ixf. 

\subsection{Shock Powered Emission}\label{subsec:Lsh}

\begin{figure*}
\centering
\subfigure{\includegraphics[width=0.328\textwidth]{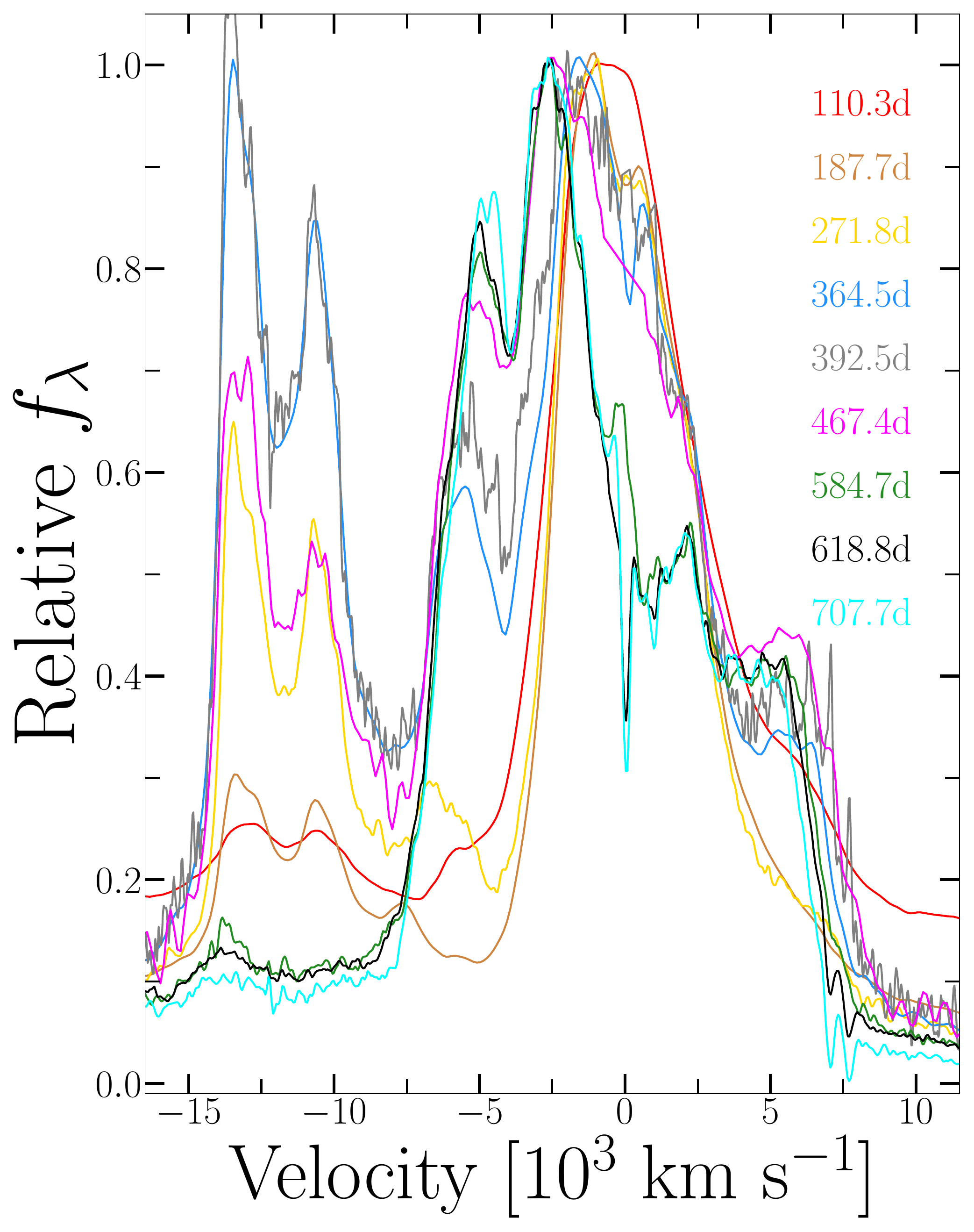}}
\subfigure{\includegraphics[width=0.33\textwidth]{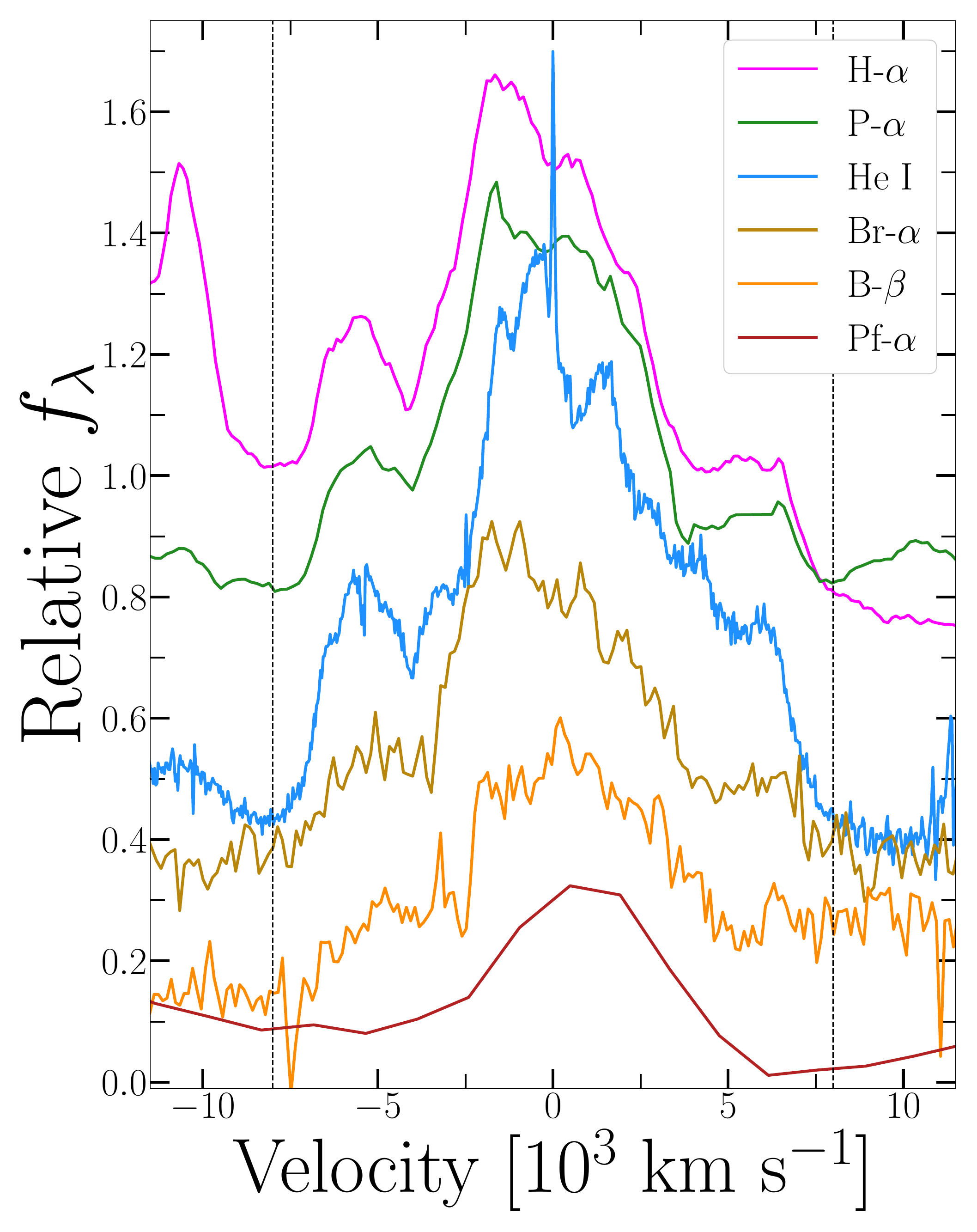}}
\subfigure{\includegraphics[width=0.33\textwidth]{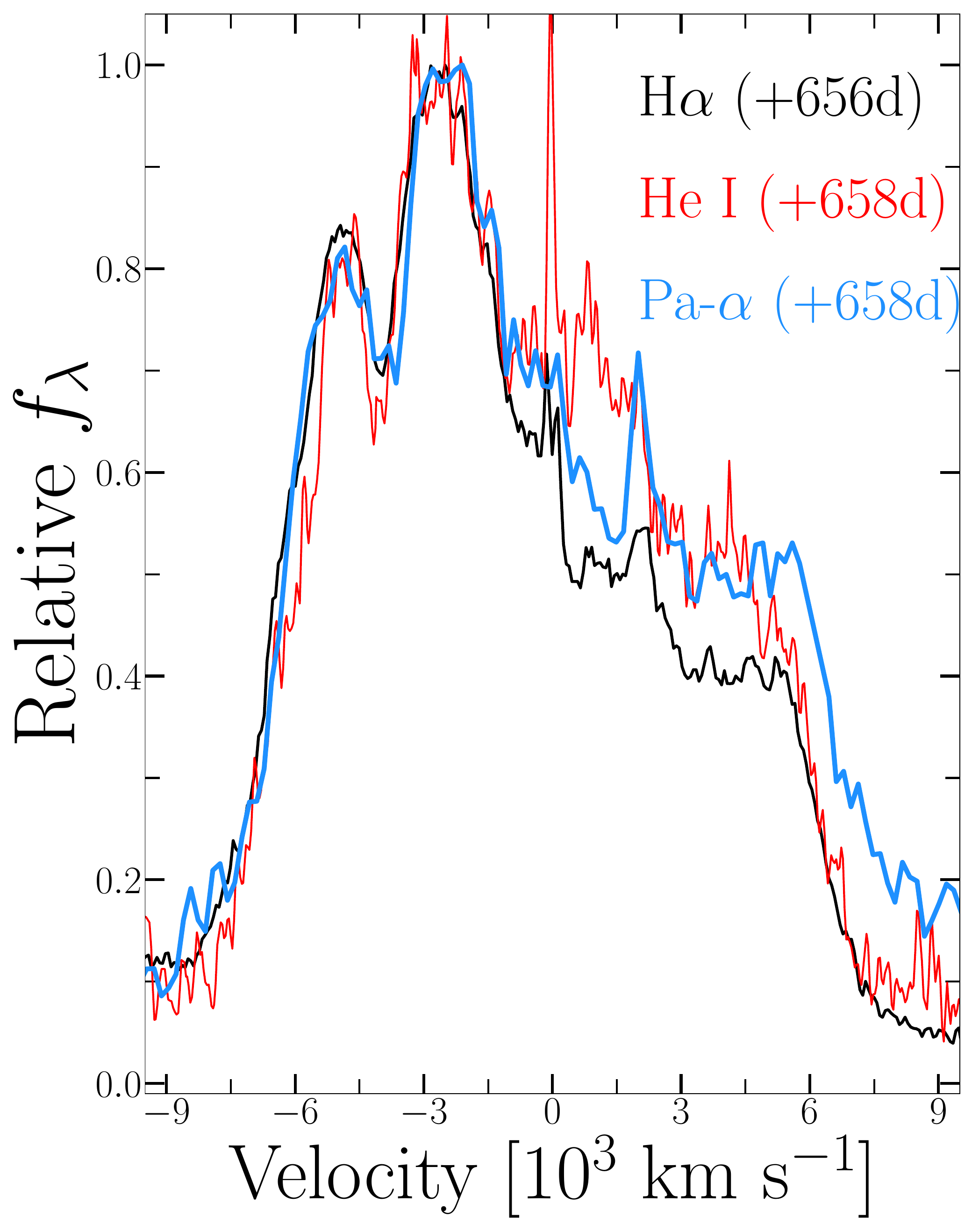}}
\caption{ {\it Left:} H$\alpha$/[\ion{O}{i}] spectral region shown in velocity space relative to H$\alpha$ rest wavelength for phases $\delta t = 110 - 708$~days. Spectra have been normalized to peak flux. H$\alpha$ shows a multi-component structure where the narrow emission line profile comes from the inner ejecta ($v < 5000~\kms$) and the ``boxy'' underlying profile comes from the CDS ($v > 5000~\kms$), both of which are heavily affected by dust attenuation. {\it Middle:} Optical-to-MIR hydrogen emission lines at $\delta t = 364 - 374$~days. Blue/red asymmetry in the narrow component of H$\alpha$ confirms dust formation in the inner ejecta, which is also supported by the wavelength dependence of such line asymmetry our to IR wavelengths. {\it Right:} H$\alpha$, Pa-$\alpha$, and \ion{He}{i} $\lambda1.083~\mu$m spectral regions in velocity space at $\delta t = 656 - 658$~days. The possible wavelength dependent blue/red asymmetry in both the ``boxy'' profile suggests the presence of dust in the CDS. \label{fig:HA} }
\end{figure*}

As shown in Figures \ref{fig:LC_all} \& \ref{fig:LC_bol}, an excess emission beyond what is expected from pure radioactive decay first becomes apparent at UV wavelengths as the {\it Swift}-UVOT light curve becomes flat at $\delta t > 200$~days and remains bright at $>0.5$~mag above the host galaxy background -- these observations are also consistent with observed emission in the near- and far-UV at $\delta t = 311 \ \& \ 619$~days \citep{Bostroem25}. Host galaxy subtracted optical and NIR photometry shows a similar flattening at $\delta t > 500$~days and the pseudobolometric light curve (e.g., Fig. \ref{fig:LC_bol}) deviates from a radioactive decay decline rate as shock powered emission from persistent SN ejecta-CSM interaction becomes the dominant power source in SN~2023ixf. Here we define ``shock powered emission'' as all additional UVOIR luminosity that does not come from absorbed radioactive decay power e.g., the reprocessed X-ray luminosity from the radiative RS. The presence of shock powered emission in SN~2023ixf is also confirmed by the spectral evolution shown in Figure \ref{fig:spec_all} as the forbidden emission lines observed at $\delta t \approx 1$~year fade and prominent boxy H$\alpha$, H$\beta$, P$\alpha$, \ion{He}{i} and [\ion{Fe}{ii}] emission become dominant at $\delta t \approx 620$~days -- this transition in H$\alpha$ is shown in Figure \ref{fig:HA}. The presence of these emission line profiles is direct evidence for a radiative reverse shock that injects power into both the CDS and the inner ejecta \citep{Nymark06, Chevalier17, dessart22}. Similarly, prominent Ly$\alpha$ and \ion{Mg}{ii}~$\lambda2800$ emission is observed at UV wavelengths, likely from the CDS (e.g., \citealt{Fransson05, Dessart23b, Bostroem24, Bostroem25}. Furthermore, evidence for ongoing CSM interaction is given by persistent X-ray and radio emission, both of which imply a mass loss rate of $\dot M \approx 10^{-4}$~\mdot\ at shock radii $>10^{15}$~cm \citep{Panjkov24, Chandra24, Nayana25}. However, this shock powered emission source from ejecta interaction with a likely steady-state wind has remained subdominant compared to radioactive decay power until $\delta t \approx 500$~days in optical/IR bands and its emergence at these late times is not the result of SN ejecta interaction with some detached shell of CSM.

%We note that the multi-wavelength evolution of SN~2023ixf confirms that shock powered emission has always been present as the SN ejecta continues to interact with the likely steady-state wind ahead of the forward shock. However, it is overall unavoidable to have some radiation produced via dissipation of kinetic energy through shocks, and this would be true in SN~2023ixf, as in all SNe, even in the absence of X-ray and radio detections. 

%Here we define ``shock powered emission'' as the emergent UVOIR luminosity from reprocessed X-ray luminosity from the radiative RS i.e., all additional UVOIR luminosity that does not come from absorbed radioactive decay power. 

\begin{figure*}
\centering
\subfigure{\includegraphics[width=0.49\textwidth]{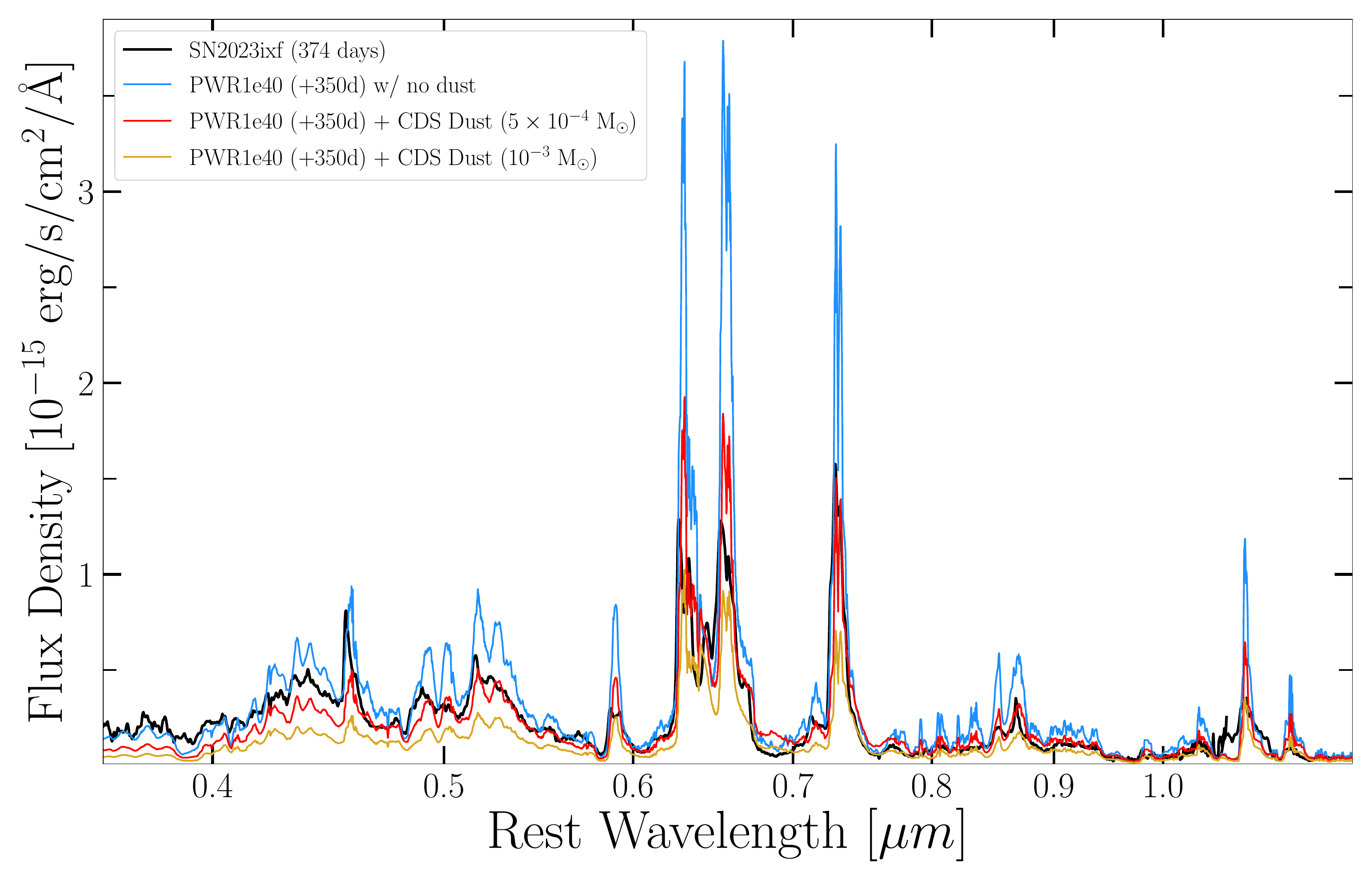}}
\subfigure{\includegraphics[width=0.49\textwidth]{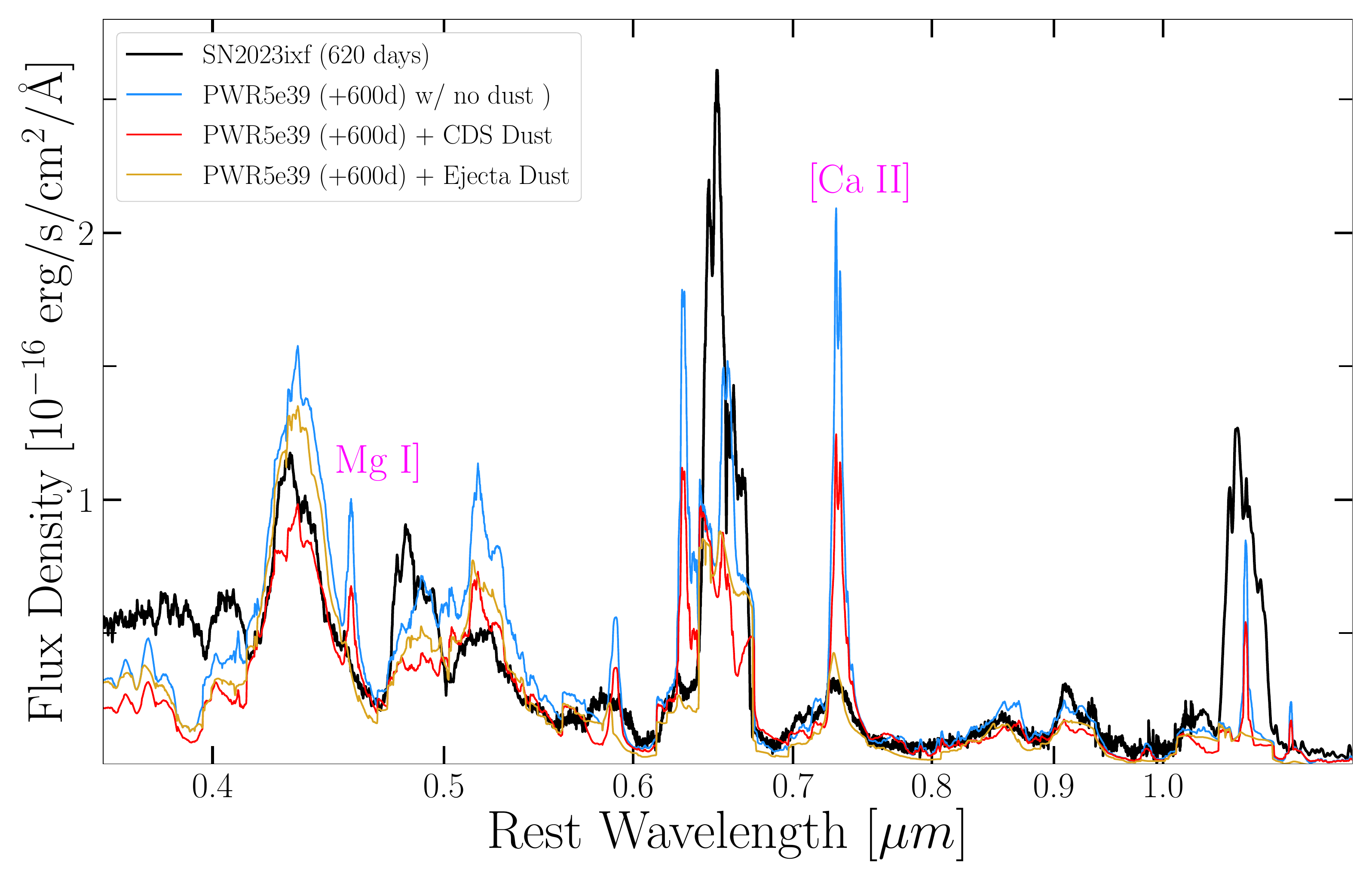}}
\caption{ {\it Left:} Best-matched {\tt CMFGEN} model spectra (red and gold lines) at $\delta t = 350$~days, with CDS dust and shock power included, compared to SN~2023ixf at $\delta t = 374$~days (black). {\tt CMFGEN} model spectra with no added dust plotted in blue shows the need for dust in the CDS to match overall SN luminosity and spectral features.  {\it Right:} Shock power {\tt CMFGEN} model spectra with at $\delta t = 600$~days with no dust (blue), CDS dust (red) and ejecta dust (gold) compared to SN~2023ixf at $\delta t = 620$~days (black). The PWR1e40 model at $\delta t = 600$~days would better match the H$\alpha$ emission but over-predicts the line emission in the UV. The disappearance of emission lines such as \ion{Mg}{i}] and [\ion{Ca}{ii}] confirms the presence of dust within the inner SN ejecta. \label{fig:cmfgenzoom} }
\end{figure*}

In order to quantify the relative contributions from shock and radioactive decay power, we construct SEDs of SN~2023ixf at $\delta t = 374 \ \& \ 620$~days, which cover a wavelength range of $0.1 - 30~\mu$m. Because the UV spectrum was obtained $\delta t = 311$~days, we scale the spectrum based on a linear interpolation of the integrated UV flux at the $+311$~day and $+619$~day {\it HST} observation epochs. At $\delta t = 374$~days, the bolometric luminosity in this wavelength range is $L_{\rm UVOIR} = (3.5 \pm 0.15)\times 10^{40}$~erg~s$^{-1}$ and the expected luminosity from radioactive decay power absorbed by the ejecta at this phase is $L_{\rm decay, abs} = (1.0 \pm 0.1)\times 10^{40}$~erg~s$^{-1}$ using the parameters derived in \S\ref{subsec:Lrad}. Consequently, the emergent shock powered emission across this wavelength range is $L_{\rm sh} = (2.6 \pm 0.15)\times 10^{40}$~erg~s$^{-1}$ at $\delta t = 374$~days. Then, at $\delta t = 620$~days, we find $L_{\rm UVOIR} = (1.4 \pm 0.10)\times 10^{40}$~erg~s$^{-1}$, $L_{\rm decay, abs} = (4.8 \pm 0.1)\times 10^{38}$~erg~s$^{-1}$, and $L_{\rm sh} = (1.3 \pm 0.10)\times 10^{40}$~erg~s$^{-1}$. Additionally, within this phase range, the observed X-ray emission (e.g., Fig. \ref{fig:Xray}) remains relatively constant with an unabsorbed 0.3-10~keV luminosity of $L_{X} \approx 10^{39}$~erg~s$^{-1}$. 

% the observed X-ray emission is arising from the FS. The emission from the RS is completely thermalized within the CDS and emitted as UV/IR/O

As discussed in \S\ref{SubSec:Xray}, the observed late-time X-ray emission is arising primarily from the FS, which is predicted to be almost completely adiabatic at this phase \citep{Fransson96, Chevalier17}. The RS X-ray emission is completely thermalized in the CDS and re-emitted as the observed shock powered emission in SN~2023ixf. For ejecta-CSM interaction with a constant mass-loss rate, the thermal bremsstrahlung (``free-free'') bolometric emission from an adiabatic shock can be described as:

\begin{equation}\label{eqn:Lff}
    L_{\rm ff} = 3\times 10^{39} \ {\rm g_{ff}} \ C_n \ \Big(\frac{\dot M_{-5}}{v_{\rm w10}}\Big)^2 \ t_{10}^{-1} \ \Big (\frac{T_e}{T_i}\Big)^{0.5} \rm erg \ s^{-1}  
\end{equation}
where $\dot M_{-5}$ is mass-loss rate in units of $10^{-5}$~\mdot, $v_{\rm w10}$ is the wind velocity in units of $10~\kms$, $\rm g_{ff}$ is the Gaunt factor (assumed to be of order unity), $T_e$ is electron temperature, $T_i$ is ion temperature, $C_n = 1$ for the FS and $C_n = (n-3)(n-4)^2 / (4(n-2)$ for the RS \citep{Chevalier03, Chevalier17}. We choose a wind velocity of $v_w = 20 \pm 5~\kms$, which has uncertainties that account for the first narrow line velocity measurement of $\sim25~\kms$ at 1.5~days \citep{Dickinson25} and any prior radiative acceleration during shock breakout \citep{Dessart25rad}. We adopt $T_e/T_i = 0.1$ based on the observed X-ray spectrum temperature and the expected ion temperature given a shock velocity of $\sim10^{4}~\kms$ as well as $n = 12$, and $\dot M = 10^{-4}$~\mdot. Using these parameters, we find expected forward shock luminosities of $L_{\rm FS} = (6.4 \pm 3.8) \times 10^{38}$~erg~s$^{-1}$ at $\delta t = 374$~days and $L_{\rm FS} = (3.8 \pm 1.9) \times 10^{38}$~erg~s$^{-1}$ at $\delta t = 620$~days. 

\begin{figure*}
\centering
\subfigure{\includegraphics[width=0.99\textwidth]{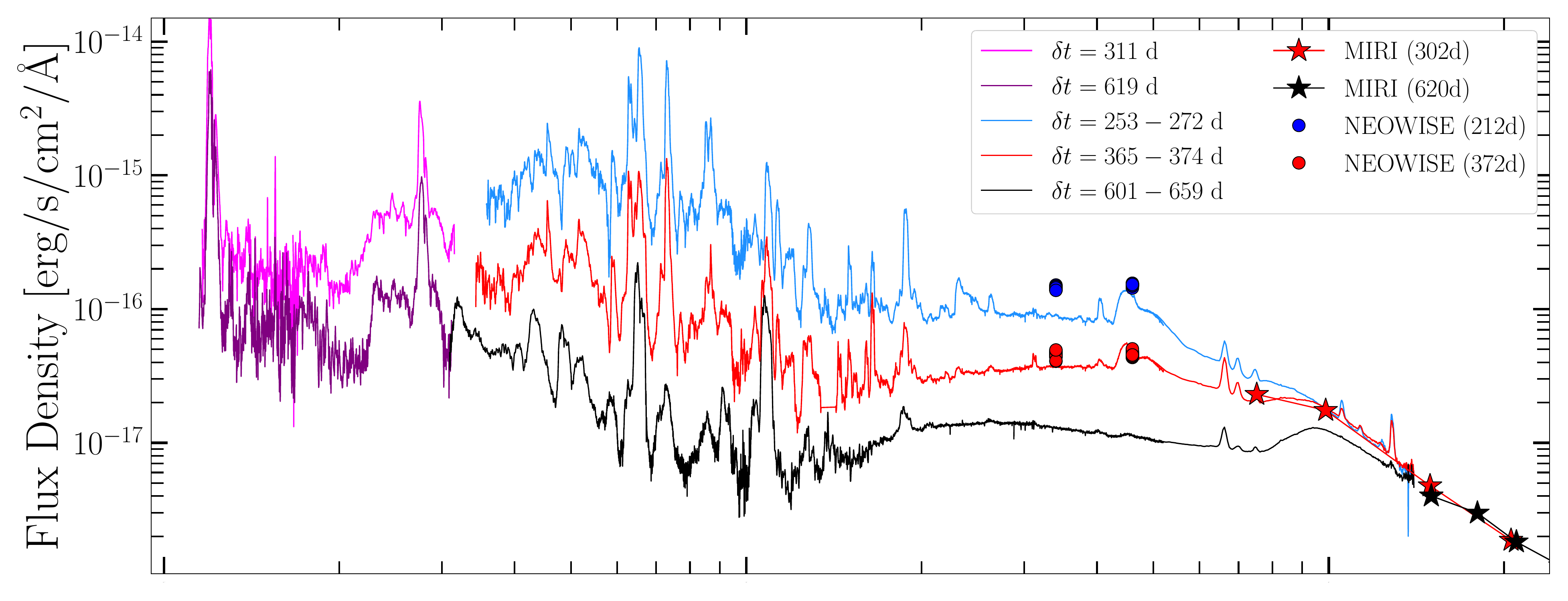}}\\
\vspace{-6mm}
\subfigure{\includegraphics[width=0.99\textwidth]{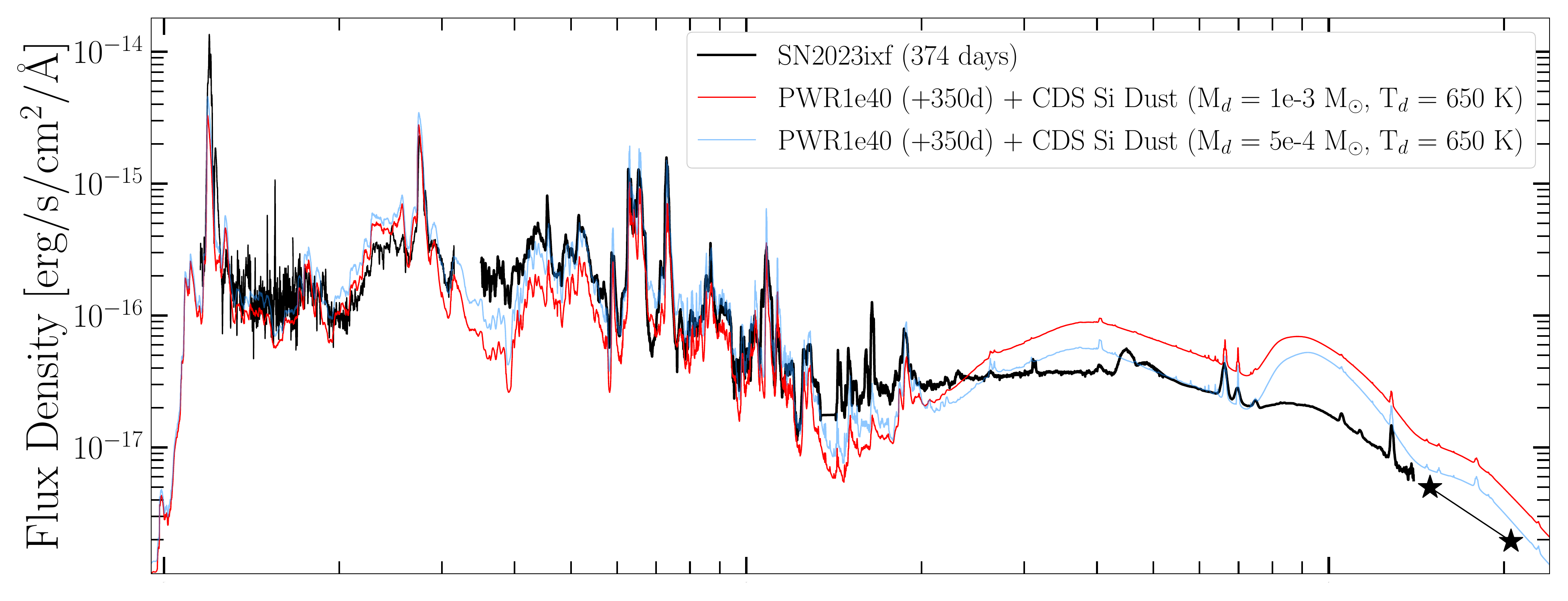}}\\
\vspace{-6mm}
\subfigure{\includegraphics[width=0.99\textwidth]{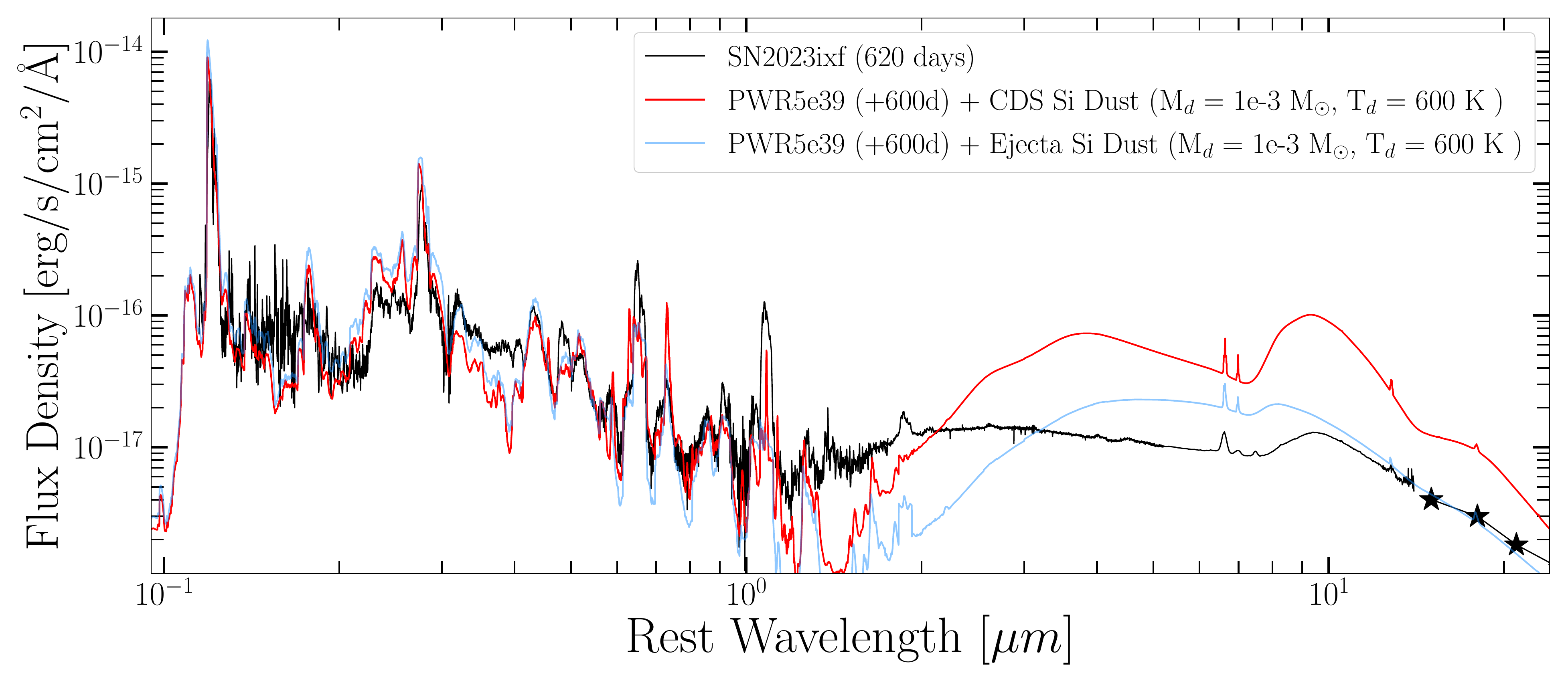}}
\caption{ {\it Top:} Multi-epoch UVOIR spectra of SN~2023ixf covering a phase range of $\delta t = 253 - 659$~days. {\it Middle/Bottom:} Best-matched {\tt CMFGEN} model spectra (red lines) at $\delta t = 350 \ \& \ 600$~days compared to UVOIR spectrum of SN~2023ixf at $\delta t = 374 \ \& \ 620$~days (black). Model spectra include $(0.5 = 1) \times 10^{40}$~erg~s$^{-1}$ of shock power and $(0.5 - 1) \times 10^{-3}~\Msun$ of silicate dust with grain size of $1~\mu$m and temperature of $600-650$~K. \label{fig:cmfgen} }
\end{figure*}

For a radiative RS, we use Eqn 3.17 from \cite{Fransson96} where the bolometric luminosity goes as:

\begin{equation}\label{eqn:LRS}
\begin{split}
    L_{\rm rev} = 1.57\times 10^{41} \frac{(2(3-s)^2(n-3)(n-4)}{(4-s)(n-s)^3} \frac{\dot M_{-5}}{v_{\rm w10}}\\ \times \ V_4^{5-s} \Big (\frac{t}{11.57 \rm \ days} \Big )^{2-s} \rm erg \ s^{-1}  
\end{split}
\end{equation}

where $V_4$ is the ejecta velocity at the RS in units of $10^4~\kms$ and $s$ is the CSM density power-law index. For a radiative cooling RS and a $s=2$ density profile, the time variable above disappears and the luminosity decreases as $L_{\rm rev} \propto t^{-(15-6s+sn-2n)/(n-s)}$, which then goes as $L_{\rm rev} \propto t^{-0.3}$ for $n=12$ and $s=2$ given the time dependence of the ejecta velocity (e.g., see Eqn. 2.2 in \citealt{Fransson96}). We adopt $\dot M = 10^{-4}$~\mdot ($v_w = 20 \pm 5~\kms$) and choose an ejecta velocity of $V_{\rm ej} = 6500 \pm 1000~\kms$, which covers the maximum velocity observed in the ``boxy'' H$\alpha$ profile (e.g., Fig. \ref{fig:HA}). Applying Eqn. 3.14 in \cite{Fransson96}, we confirm that the RS is indeed radiative because the cooling time is much less than the age of the SN at both epochs e.g., $t_{\rm cool} = 22$~days at $\delta t = 374$~days and $t_{\rm cool} = 64$~days at $\delta t = 620$~days. Using these parameters and the equation above, we find $L_{\rm RS} = (2.6 \pm 1.4) \times 10^{40}$~erg~s$^{-1}$ at $\delta t = 374$~days and $L_{\rm RS} = (2.3 \pm 1.2) \times 10^{40}$~erg~s$^{-1}$ at $\delta t = 620$~days. At these phases, the shock powered emission ($L_{\rm sh}$) observed in SN~2023ixf is $(2.6 \pm 0.15)\times 10^{40}$~erg~s$^{-1}$ and $(1.3 \pm 0.10)\times 10^{40}$~erg~s$^{-1}$, both of which are consistent with the predicted RS emission. Overall, the emergent shock powered emission observed in SN~2023ixf indicates nearly complete thermalization of RS X-rays converted to UVOIR emission by the CDS and inner ejecta. Intriguingly, the power budget in SN~2023ixf at $\delta t = 620$~days implies that the luminous near- and mid-IR emission from silicate and carbonaceous dust grains (e.g., Figures \ref{fig:SED_all} \& \ref{fig:cmfgen}) is a product of shock powered emission since radioactive decay power is insufficient to heat the dust formed in the CDS and inner ejecta. 

% Using $\delta t = 370$~d, $v_w = 25~\kms$, $\dot M = 10^{-4}$~\mdot, $T_e/T_i = 0.1$, and $n=12$:
% \begin{itemize}
%     \item $L_{\rm FS} = 4\times 10^{38}$~erg~s$^{-1}$
%     \item $L_{\rm RS} = 6\times 10^{39}$~erg~s$^{-1}$
%     \item $L_{\rm x,obs} = 1\times 10^{39}$~erg~s$^{-1}$
%     \item $L_{\rm UVOIR} = 3\times 10^{40}$~erg~s$^{-1}$
%     \item $L_{\rm Ni-56} = 1\times 10^{40}$~erg~s$^{-1}$
% \end{itemize}

\subsection{Radiative Transfer Model of the Low-Energy Radiation}\label{subsec:cmfgen}

The multi-epoch, late-time SED of SN~2023ixf enables an unprecedented opportunity to model the UV-to-IR emission of a CSM-interacting SN~II. To do this, we compute late-time synthetic spectra with the non-LTE radiative transfer code {\tt CMFGEN} \citep{hillier12} and employ a combination of SN~II nebular models from \cite{Dessart21}, SN~II late-time shock power models from \cite{dessart22, Dessart23b}, and SN~II models in which the radiative transfer treats dust absorption, scattering and emission \citep{dessart25Dust}. Unlike models that focus on blue/red asymmetry within individual emission lines (e.g., H$\alpha$), these models account for dust absorption and thermal emission across the complete SED, but require dust location/temperature as input parameters in the \cmfgen\ calculation (i.e., not computed ab initio; see \citealt{dessart25Dust} for details). We adopt a progenitor ZAMS mass of $15.2~\Msun$, which contains a similar $^{56}$Ni mass to SN~2023ixf. However, in order to account for the greater $\gamma$-ray escape observed in SN~2023ixf and to match the overall continuum level at late-times, we scale the absorbed radioactive decay power to 40\% of its original model value. We explore a range of shock powers ($10^{39} - 10^{41}$~erg~s$^{-1}$), dust masses ($10^{-5} - 10^{-2}~\Msun$), dust temperatures ($600 - 800$~K), dust compositions (silicate vs carbonaceous), dust grain size ($0.1 - 1~\mu$m), and dust location (inner ejecta at $<2500~\kms$ or within the narrow CDS at $8000~\kms$) -- a subset of this model exploration is presented in Appendix Figure \ref{fig:cmfgenall}. We find that $\sim 10^{-3}~\Msun$ of 1$\mu$m silicate dust with $T = 600$~K yields a satisfactory match to the IR spectrum of SN~2023ixf, which shows clear emission from silicate dust at $\sim 10~\mu$m. We note that the \cmfgen\ model includes gas (atom and ions) and dust, but ignores molecules. 

\begin{deluxetable*}{ccccccc}[t!]
\tablecaption{Spectral Energy Distribution Components \label{tab:SED}}
\tablecolumns{7}
%\tablenum{2}
\tablewidth{0.45\textwidth}
\tablehead{\colhead{Phase} & \colhead{$L_{\rm UVOIR}^a$} &
\colhead{$L_{\rm decay,abs}^b$} &
\colhead{$L_{\rm sh}$} & \colhead{$L_{\rm x-ray}^c$} & \colhead{$L_{\rm FS}^d$} & \colhead{$L_{\rm RS}^e$}\\
\colhead{(days)} & \colhead{(erg~s$^{-1}$)} & \colhead{(erg~s$^{-1}$)} & \colhead{(erg~s$^{-1}$)} & \colhead{(erg~s$^{-1}$)} & \colhead{(erg~s$^{-1}$)} & \colhead{(erg~s$^{-1}$)}
}
\startdata
374 & $(3.5 \pm 0.14)\times 10^{40}$ & $(1.0 \pm 0.1)\times 10^{40}$ & $(2.6 \pm 0.15)\times 10^{40}$ & $(7.0 - 9.4)\times 10^{38}$ & $(6.4 \pm 3.2)\times 10^{38}$ & $(2.6 \pm 1.4)\times 10^{40}$ \\
620 & $(1.4 \pm 0.10)\times 10^{40}$ & $(4.8 \pm 0.1)\times 10^{38}$ & $(1.3 \pm 0.10)\times 10^{40}$ & $(7.0 - 9.4)\times 10^{38}$ & $(3.8 \pm 1.9)\times 10^{38}$ & $(2.3 \pm 1.2)\times 10^{40}$ \\
\enddata
\tablenotetext{a}{0.1-30~$\mu$m}
\tablenotetext{b}{$M(^{56} \rm Ni) = 0.058 \pm 0.0010~\Msun$, $t_{\gamma} = 264.6 \pm 2.5$~days}
\tablenotetext{c}{Unabsorbed 0.3-10~keV luminosities from FS-only model at $\delta t = 186.2 - 585.2$~days.}
\tablenotetext{d}{$\dot M = 10^{-4}$~\mdot, $v_w = 20\pm 5~\kms$, $n = 12$, $T_e/T_i = 0.1$}
\tablenotetext{e}{$\dot M = 10^{-4}$~\mdot, $v_w = 20\pm 5~\kms$, $n = 12$, $s = 2$, $V_{\rm ej} = 6500 \pm 1000~\kms$}
%\tablecomments{}
\end{deluxetable*}

As shown in Figures \ref{fig:cmfgenzoom} and \ref{fig:cmfgenall}, the inclusion of dust can also produce similar blue/red asymmetries that are observed in both the boxy and narrow components of H$\alpha$, indicative of dust formation in the inner ejecta during the first complete spectral epoch ($\delta t = 374$~day) as well as in the CDS by the next epoch ($\delta t = 620$~days). Furthermore, the asymmetries in the line profiles may also be wavelength dependent (e.g., see Fig. \ref{fig:HA}), which can be used to trace the dust opacity as has been done for type IIn SNe (e.g., SN 2010jl; \citealt{Gall14}) and could also suggest larger grain sizes of $>1~\mu$m. While some pre-existing dust external to the FS may exist, the overall CSM dust mass must be quite small because it would cause significant attenuation at UV wavelengths, which is not observed in the {\it HST} spectra. This overall evolution of dust formation is supported by the detection of molecular CO emission beginning at $\delta t > 200$~days \citep{Park25, DerKacy25} and mid-infrared silicate emission \citep{Medler25}. As shown in Appendix Figure \ref{fig:dust}, we fit both epochs of IR spectra with analytic formalisms for dust emission (e.g., see \citealt{fox10, fox11, Tinyanont19, Shahbandeh23, Pearson25, Tinyanont25})\footnote{\url{https://github.com/stinyanont/sed_et_al}} after masking out all emission lines and find consistent silicate dust masses to what is adopted in the best-matched {\tt CMFGEN} spectra. However, we find that the IR spectrum of SN~2023ixf requires an additional C-rich dust components to match the NIR emission, which is not included in the best-matched {\tt CMFGEN} model spectra with Si-rich dust. Furthermore, it is clear from the analytic model fits that the emission at $>15~\mu$m is not well-matched -- this could be due to a number of effects e.g., dust opacities, clumping, etc. \citep{Sarangi22, dessart25Dust}. Future simulations will explore combinations of various dust compositions, temperatures and structures. 

%this Si dust mass can also reproduce the IR spectra and photometry when using analytic formalisms for dust emission (e.g., see \citealt{fox10, fox11, Shahbandeh23}).

In Figure \ref{fig:cmfgen}, we present the best-matched {\tt CMFGEN} models at $\delta t = 350 \ \& \ 600$~days compared to the UVOIR SED of SN~2023ixf at $\delta t = 374 \ \& \ 620$~days. We find that a shock power of $10^{40}$~erg~s$^{-1}$ is sufficient to match the total UV emission and the most prominent line profiles such as Ly$\alpha$ and \ion{Mg}{ii}~$\lambda2800$ as well as the boxy emission observed in H$\alpha$ at $\delta t = 374$~days. At $\delta t = 600$~days, we find that the model shock power needs to be reduced to $5\times10^{39}$~erg~s$^{-1}$ so as to not over-predict the UV emission lines, but also observe that the $10^{40}$~erg~s$^{-1}$ shock power model is a better match to the optical and NIR spectra at this phase. While the reduction in total UV flux between epochs is expected from a declining RS X-ray luminosity (e.g., Eqn. \ref{eqn:Lff}), we note that the observed shock powered emission is larger than the model prediction at $\delta t = 600$~days. Furthermore, at $\delta t = 600$~days, one {\tt CMFGEN} model includes dust in the CDS and another with dust in the inner ejecta ($v < 2500~\kms)$, both of which have the most dramatic effect on the H and He emission line profiles. However, these current {\tt CMFGEN} models show inconsistency with the spectral shape and overall flux in the IR, which is likely the product of a fixed dust temperature and single dust emission location within the SN. Further analysis including hybrid dust locations and dust temperature profiles should be explored but is beyond the scope of this paper. 

% $L_{\rm RS} = (2.6 \pm 1.4) \times 10^{40}$~erg~s$^{-1}$ at $\delta t = 374$~days and $L_{\rm RS} = (2.3 \pm 1.2) \times 10^{40}$~erg~s$^{-1}$ at $\delta t = 620$~days
% \section{Discussion} \label{sec:discussion}

\section{Conclusions} \label{sec:conclusion}

In this paper, we present late-time, multi-wavelength observations of the CSM-interacting SN~II 2023ixf spanning X-ray, UV, optical, IR wavelengths. Below we summarize the primary observational findings from this work.  

\begin{itemize}
    \item Following the end of the light curve plateau, the multi-band light curve of SN~2023ixf follows a decline rate dictated by absorbed radioactive decay power. Then, at $\delta t \approx 200$~days for UV bands and $\delta t \approx 500$~days for optical/IR bands, the SN~2023ixf light curve flattens as shock powered emission ($L_{\rm sh}$) from enduring CSM-interaction dominates over absorbed radioactive decay power ($L_{\rm decay,abs}$). We model the $UBgVrizyJHK$ band pseudobolometric light curve at $\delta t < 400$~days and find $M(^{56} \rm Ni) = 0.0580 \pm 0.0010~\Msun$, $t_{\gamma} = 264.6 \pm 2.5$~days.

    \item At late-time phases, H$\alpha$ shows broad, ``boxy'' emission from a CDS with velocity of $\sim 8500~\kms$ -- this structure also being seen in UV (e.g., \ion{Mg}{ii}~$\lambda2800$, Ly-$\alpha$) and NIR (e.g., \ion{He}{i}, Pa-$\alpha$) emission lines. We interpret these observations as a transition to a regime dominated by shock powered emission where the UV/optical/NIR spectra are dominated by emission from the CDS and inner ejecta. Significant blue-red asymmetry observed in H$\alpha$, combined with luminous near- and mid-IR emission, indicates dust formation in the inner ejecta beginning at $\delta t \gtrsim 300$~days and then in the CDS by at least $\delta t \approx 620$~days.
    
    \item We quantify the relative contributions from absorbed radioactive decay power and shock powered emission in SN~2023ixf by constructing $0.1 - 30~\mu$m SEDs at $\delta t = 374 \ \& \ 620$~days. In the first epoch, we find $L_{\rm UVOIR} = (3.5 \pm 0.15)\times 10^{40}$~erg~s$^{-1}$ and $L_{\rm decay,abs} = (1.0 \pm 0.1)\times 10^{40}$~erg~s$^{-1}$, which results in an emergent shock powered emission of $L_{\rm sh} = (2.6 \pm 0.15)\times 10^{40}$~erg~s$^{-1}$. Then, at $\delta t = 620$~days, we find $L_{\rm UVOIR} = (1.4 \pm 0.10)\times 10^{40}$~erg~s$^{-1}$, $L_{\rm decay,abs} = (4.8 \pm 0.1)\times 10^{38}$~erg~s$^{-1}$, and $L_{\rm sh} = (1.3 \pm 0.10)\times 10^{40}$~erg~s$^{-1}$. 

    \item We estimate unabsorbed 0.3-10~keV luminosities at $\delta t = 186.2 - 646.3$~days by modeling the {\it Swift}-XRT spectra with an absorbed thermal bremsstrahlung model. Despite indirect evidence for the presence of a radiative RS from UV-to-IR spectra and the late-time flattening of the X-ray light curve, we find no statistical evidence for a secondary, lower temperature component from this shock in the X-ray spectra. We use the emission measure from the adiabatic FS component to calculate unshocked CSM densities at shock radii of $10^{16} - 10^{17}$~cm and confirm that the CSM density profile continues to be well-described by a constant mass loss rate of $\dot M = 10^{-4}$~\mdot\ ($v_w = 20~\kms$).

    \item We find that the measured shock powered emission contribution to the $0.1 - 30~\mu$m SED at $\delta t = 374 \ \& \ 620$~days can be explained by nearly complete thermalization of power from a radiative RS into both the CDS and inner SN ejecta. Using the analytic formalism for thermal bremsstrahlung emission from ejecta-CSM interaction, the observed shock powered emission in SN~2023ixf can be reproduced by RS emission from a constant mass-loss rate of $\dot M = 10^{-4}$~\mdot, assuming $v_w = 20 \pm 5~\kms$, $s = 2$ and $n = 12$ (Eqn. \ref{eqn:LRS}). Furthermore, these same parameters predict a FS luminosity that is consistent with the observed X-ray light curve. 

    \item We construct \cmfgen\ model spectra at $\delta t = 350 \ \& \ 600$~days, which include absorbed radioactive decay power, shock power and dust emission. As shown in Figure \ref{fig:SED_all}, model spectra with $L_{\rm sh} = (0.5 - 1) \times 10^{40}$erg~s$^{-1}$ and $(0.5 - 1)\times 10^{-3}~\Msun$ of silicate dust can reproduce the global properties of the late-time SN~2023ixf $0.1 - 30~\mu$m SEDs.  

\end{itemize}

Given its present evolution, SN~2023ixf represents an ideal laboratory to study shock powered emission through continued, multi-wavelength observations. Future X-ray spectra should be able to confirm multiple temperature components from the emerging RS and also constrain the CSM densities at shock radii $>10^{17}$~cm. These observations, coupled with complete UVOIR SED coverage will enable more robust constraints on the thermalization efficiency of the RS luminosity and the mechanisms for heating newly formed dust.

\facilities{{\it Swift} UVOT/XRT, Shane Telescope (Kast), Las Cumbres Observatory, Lulin Observatory, Thacher Observatory, Keck Observatory (LRIS/NIRES/KCWI), Hale Telescope (WIRC), Zwicky Transient Facility (ZTF), {\it JWST} (NIRSpec/MIRI), {\it HST} (STIS)}

\software{IRAF (Tody 1986, Tody 1993), photpipe \citep{Rest+05}, DoPhot \citep{Schechter+93}, HOTPANTS \citep{becker15}, YSE-PZ \citep{Coulter22, Coulter23}, \cmfgen\ \citep{hillier12, dessart22}, Lpipe \citep{Perley19}, HEAsoft \citep{HEAsoft} }

\section{Acknowledgments} \label{Sec:ack}

W.J.-G.\ is supported by NASA through Hubble Fellowship grant HSTHF2-51558.001-A awarded by the Space Telescope Science Institute, which is operated for NASA by the Association of Universities for Research in Astronomy, Inc., under contract NAS5-26555.
%Who is this?
%This research was supported in part by the NSF under grant PHY-1748958.
% Dessart? yes
This work was granted access to the HPC resources of TGCC under the allocation 2024 -- A0170410554 made by GENCI, France.
C.D.K.\ gratefully acknowledges support from the NSF through AST-2432037, the HST Guest Observer Program through HST-SNAP-17070 and HST-GO-17706, and from JWST Archival Research through JWST-AR-6241 and JWST-AR-5441.
% Auchettl? yes
Parts of this research were supported by the Australian Research Council Centre of Excellence for Gravitational Wave Discovery (OzGrav), through project number CE230100016.
The Margutti team at UC Berkeley is partially funded by the Heising-Simons Foundation under grants \#2018-0911 and \#2021-3248 (PI R. Margutti).
% The H-S grant is listed twice (both above and below)
The TReX team at UC Berkeley is partially supported by the NSF under grant AST-2224255, and by the Heising-Simons Foundation under grant \#2021-3248 (PI R. Margutti).
V.V.D.'s X-ray astronomy research is currently supported by Chandra Grants G03-24041X and TM4-25004X.
K.A.B.\ is supported by an LSSTC Catalyst Fellowship; this publication was thus made possible through the support of grant 62192 from the John Templeton Foundation to LSSTC. The opinions expressed in this publication are those of the authors and do not necessarily reflect the views of LSSTC or the John Templeton Foundation.
R.C.\ acknowledges support from NASA {\it Swift} grant 80NSSC22K0946.
The UCSC team is supported in part by NASA grants 80NSSC23K0301 and 80NSSC24K1411; and a fellowship from the David and Lucile Packard Foundation to R.J.F.
% Is this right? yes (WJG)
T.A.\ acknowledges generous support from the David and Lucile Packard Foundation.
C.G.\ is supported by a VILLUM FONDEN Young Investigator Grant (VIL25501) and Experiment Grant (VIL69896).
C.L.\ acknowledges support under DOE award DE-SC0010008 to Rutgers University.
G.N.\ is funded by NSF CAREER grant AST-2239364, supported in-part by funding from Charles Simonyi. G.N.\ also gratefully acknowledges NSF support from AST-2206195, OAC-2311355, AST-2432428, as well as AST-2421845 and funding from the Simons Foundation for the NSF-Simons SkAI Institute. G.N.\ is also supported by the DOE through the Department of Physics at the University of Illinois, Urbana-Champaign (\# 13771275), and support from the HST Guest Observer Program through HST-GO-17128 (PI: R.\ Foley).
S.-H.P.\ was supported by the National Research Foundation of Korea (NRF) NRF-2019R1A2C2010885 and NRF-2022H1D3A2A01096434.

%MAST

Some/all of the data presented in this paper were obtained from the Mikulski Archive for Space Telescopes (MAST) at the Space Telescope Science Institute. The specific observations analyzed can be accessed via \dataset[https://doi.org/10.17909/aj20-5b11]{https://doi.org/10.17909/aj20-5b11}. STScI is operated by the Association of Universities for Research in Astronomy, Inc., under NASA contract NAS5–26555. Support to MAST for these data is provided by the NASA Office of Space Science via grant NAG5–7584 and by other grants and contracts.

%HST + JWST
This research is based on observations made with the NASA/ESA Hubble Space Telescope obtained from the Space Telescope Science Institute, which is operated by the Association of Universities for Research in Astronomy, Inc., under NASA contract NAS 5–26555. These observations are associated with programs HST-GO-17497 (PI Valenti) and HST-GO-17772 (PI Bostroem).

This work is based in part on observations made with the NASA/ESA/CSA James Webb Space Telescope. The data were obtained from the Mikulski Archive for Space Telescopes at the Space Telescope Science Institute, which is operated by the Association of Universities for Research in Astronomy, Inc., under NASA contract NAS 5-03127 for JWST. These observations are associated with programs JWST-GO-3921 (PI Fox), JWST-DD-4575 (PI Ashall), and JWST-GO-5290 (PI Ashall). The authors acknowledge the teams led by PI Ashall and PI Fox for developing their observing program with a zero-exclusive-access period.

%Swift
We acknowledge the use of public data from the {\it Swift} data archive.

% YSE
The Young Supernova Experiment (YSE) and its research infrastructure is supported by the European Research Council under the European Union's Horizon 2020 research and innovation programme (ERC Grant Agreement 101002652, PI K.\ Mandel), the Heising-Simons Foundation (2018-0913, PI R.\ Foley; 2018-0911, PI R.\ Margutti), NASA (NNG17PX03C, PI R.\ Foley), NSF (AST--1720756, AST--1815935, PI R.\ Foley; AST--1909796, AST-1944985, PI R.\ Margutti), the David \& Lucille Packard Foundation (PI R.\ Foley), VILLUM FONDEN (project 16599, PI J.\ Hjorth), and the Center for AstroPhysical Surveys (CAPS) at the National Center for Supercomputing Applications (NCSA) and the University of Illinois Urbana-Champaign.
% Pan-STARRS
Pan-STARRS is a project of the Institute for Astronomy of the University of Hawaii, and is supported by the NASA SSO Near Earth Observation Program under grants 80NSSC18K0971, NNX14AM74G, NNX12AR65G, NNX13AQ47G, NNX08AR22G, 80NSSC21K1572, and by the State of Hawaii.  The Pan-STARRS1 Surveys (PS1) and the PS1 public science archive have been made possible through contributions by the Institute for Astronomy, the University of Hawaii, the Pan-STARRS Project Office, the Max-Planck Society and its participating institutes, the Max Planck Institute for Astronomy, Heidelberg and the Max Planck Institute for Extraterrestrial Physics, Garching, The Johns Hopkins University, Durham University, the University of Edinburgh, the Queen's University Belfast, the Harvard-Smithsonian Center for Astrophysics, the Las Cumbres Observatory Global Telescope Network Incorporated, the National Central University of Taiwan, STScI, NASA under grant NNX08AR22G issued through the Planetary Science Division of the NASA Science Mission Directorate, NSF grant AST-1238877, the University of Maryland, Eotvos Lorand University (ELTE), the Los Alamos National Laboratory, and the Gordon and Betty Moore Foundation.

% Keck
Some of the data presented herein were obtained at Keck Observatory, which is a private 501(c)3 non-profit organization operated as a scientific partnership among the California Institute of Technology, the University of California, and the National Aeronautics and Space Administration. The Observatory was made possible by the generous financial support of the W.\ M.\ Keck Foundation.
The authors wish to recognize and acknowledge the very significant cultural role and reverence that the summit of Maunakea has always had within the indigenous Hawaiian community.  We are most fortunate to have the opportunity to conduct observations from this mountain.

% Kast
A major upgrade of the Kast spectrograph on the Shane 3~m telescope at Lick Observatory was made possible through generous gifts from the Heising-Simons Foundation as well as William and Marina Kast. Research at Lick Observatory is partially supported by a generous gift from Google.

% ZTF
Based on observations obtained with the Samuel Oschin Telescope 48-inch and the 60-inch Telescope at the Palomar Observatory as part of the Zwicky Transient Facility project. ZTF is supported by the National Science Foundation under Grant Nos. AST-1440341, AST-2034437, and a collaboration including Caltech, IPAC, the Weizmann Institute for Science, the Oskar Klein Center at Stockholm University, the University of Maryland, the University of Washington, Deutsches Elektronen-Synchrotron and Humboldt University, the TANGO Consortium of Taiwan, the University of Wisconsin at Milwaukee, Trinity College Dublin, Lawrence Livermore National Laboratories, and IN2P3, France. Operations are conducted by COO, IPAC, and UW.

This publication has made use of data collected at Lulin Observatory, partly supported by the TAOvA with the NSTC grant 113-2740-M-008-005. 

% YSE-PZ
YSE-PZ was developed by the UC Santa Cruz Transients Team, supported in part by NASA grants NNG17PX03C, 80NSSC19K1386, and 80NSSC20K0953; NSF grants AST--1518052, AST--1815935, and AST--1911206; the Gordon \& Betty Moore Foundation; the Heising-Simons Foundation; a fellowship from the David and Lucile Packard Foundation to R.\ J.\ Foley; Gordon and Betty Moore Foundation postdoctoral fellowships and a NASA Einstein fellowship, as administered through the NASA Hubble Fellowship program and grant HST-HF2-51462.001, to D.\ O.\ Jones; and a National Science Foundation Graduate Research Fellowship, administered through grant No.\ DGE-1339067, to D.\ A.\ Coulter.

%%%%%%%%%%%%%%%
\bibliographystyle{aasjournal} 
\bibliography{references} 

%%%%%%%%%%%%

\clearpage
\appendix

\begin{figure*}
\centering
\subfigure{\includegraphics[width=0.33\textwidth]{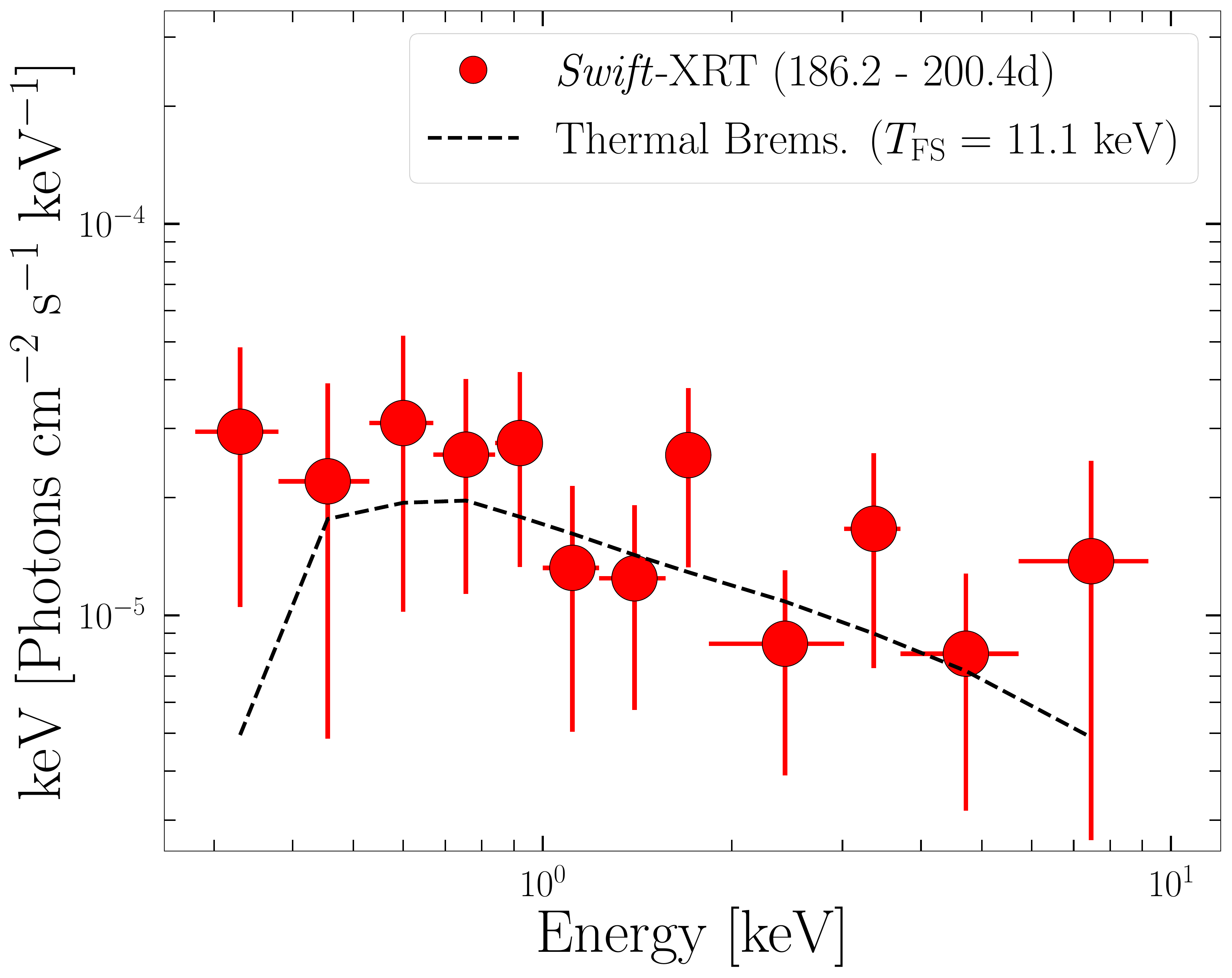}}
\subfigure{\includegraphics[width=0.33\textwidth]{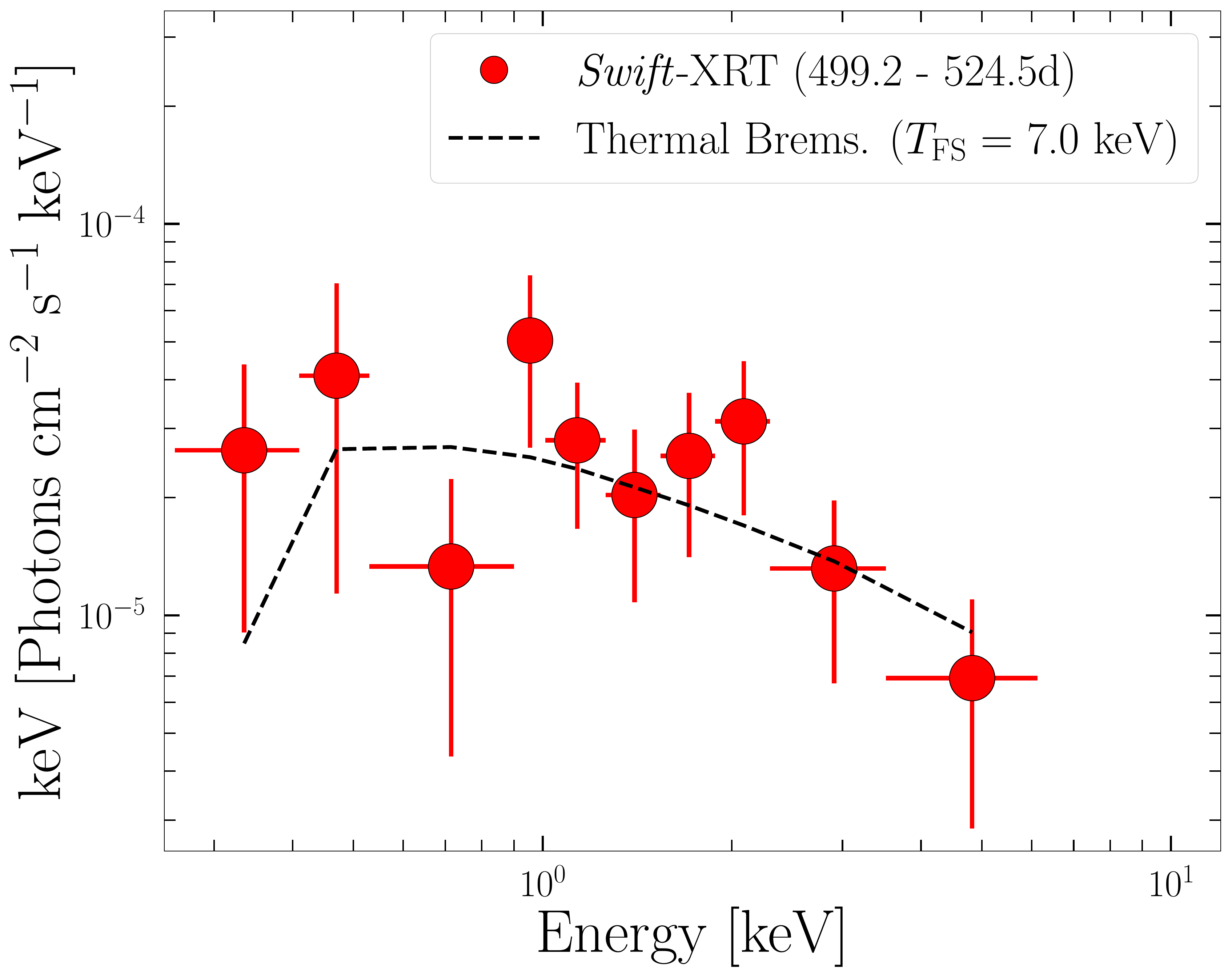}}
\subfigure{\includegraphics[width=0.33\textwidth]{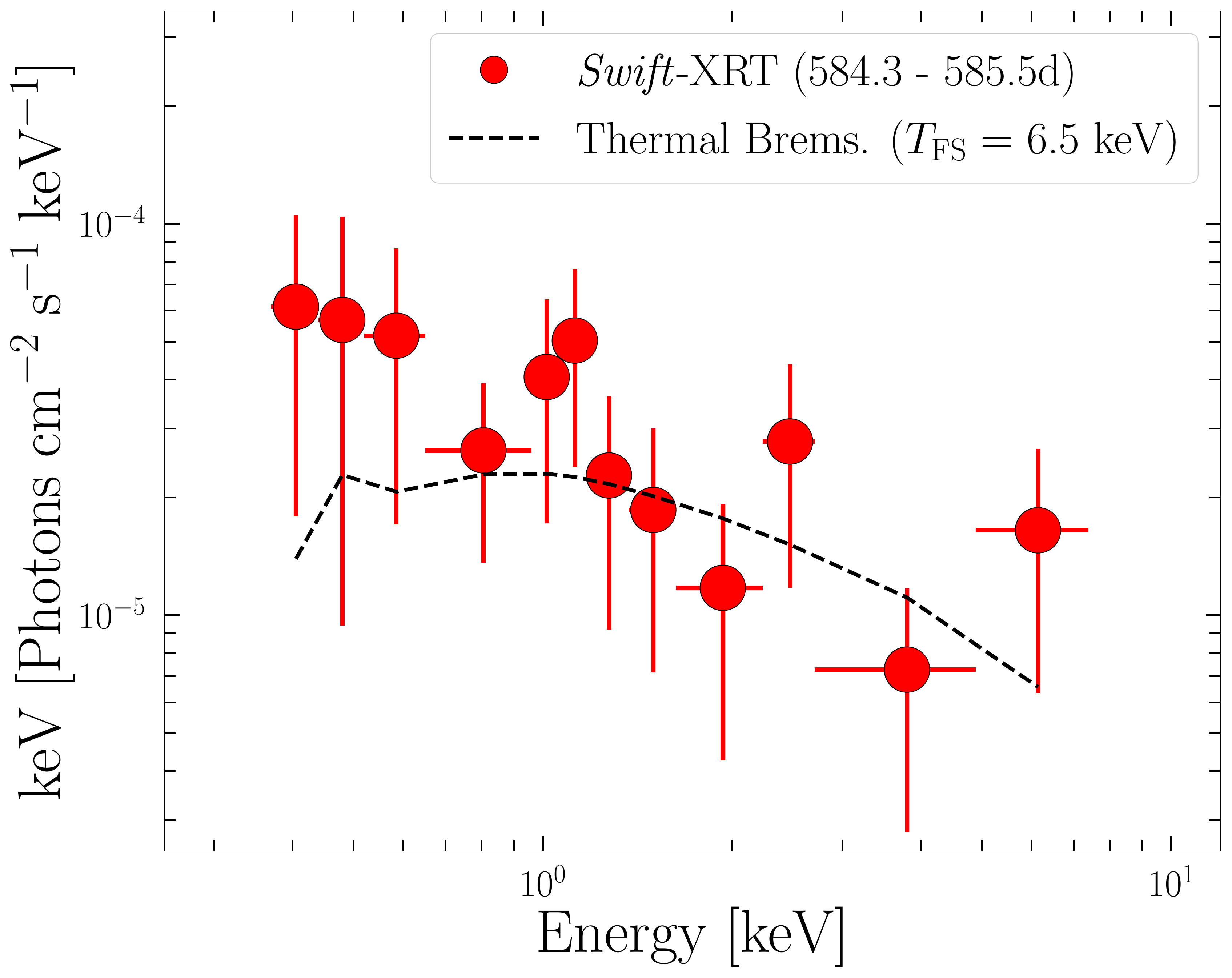}}
\caption{ Single temperature, thermal bremsstrahlung model fit (black dashed lines) to three late-time epochs of {\it Swift}-XRT data (red circles). Temperature is fixed in these models following the decline rate derived by \cite{Nayana25}. \label{fig:XRT} }
\end{figure*}

\begin{figure*}
\centering
\subfigure{\includegraphics[width=0.49\textwidth]{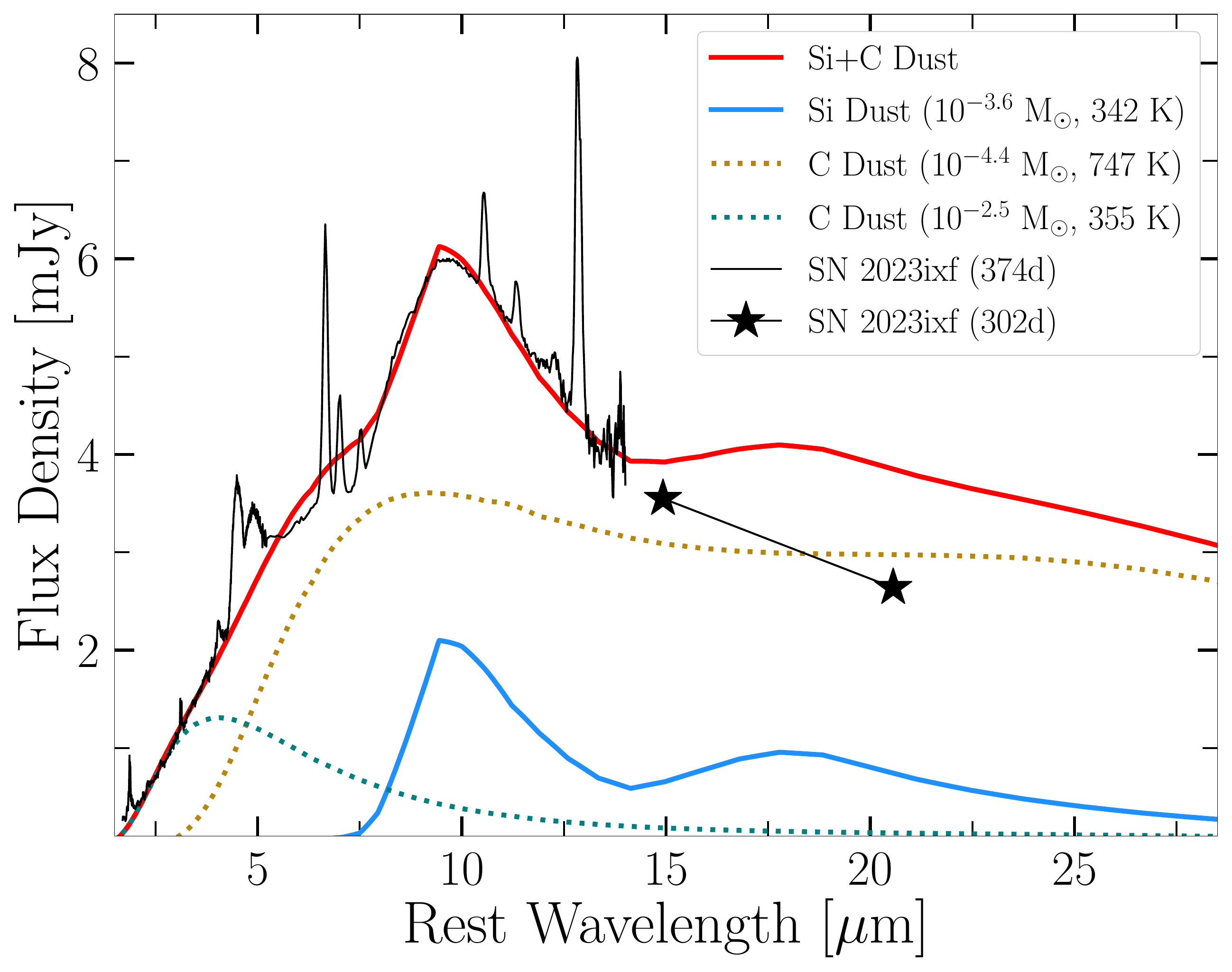}}
\subfigure{\includegraphics[width=0.49\textwidth]{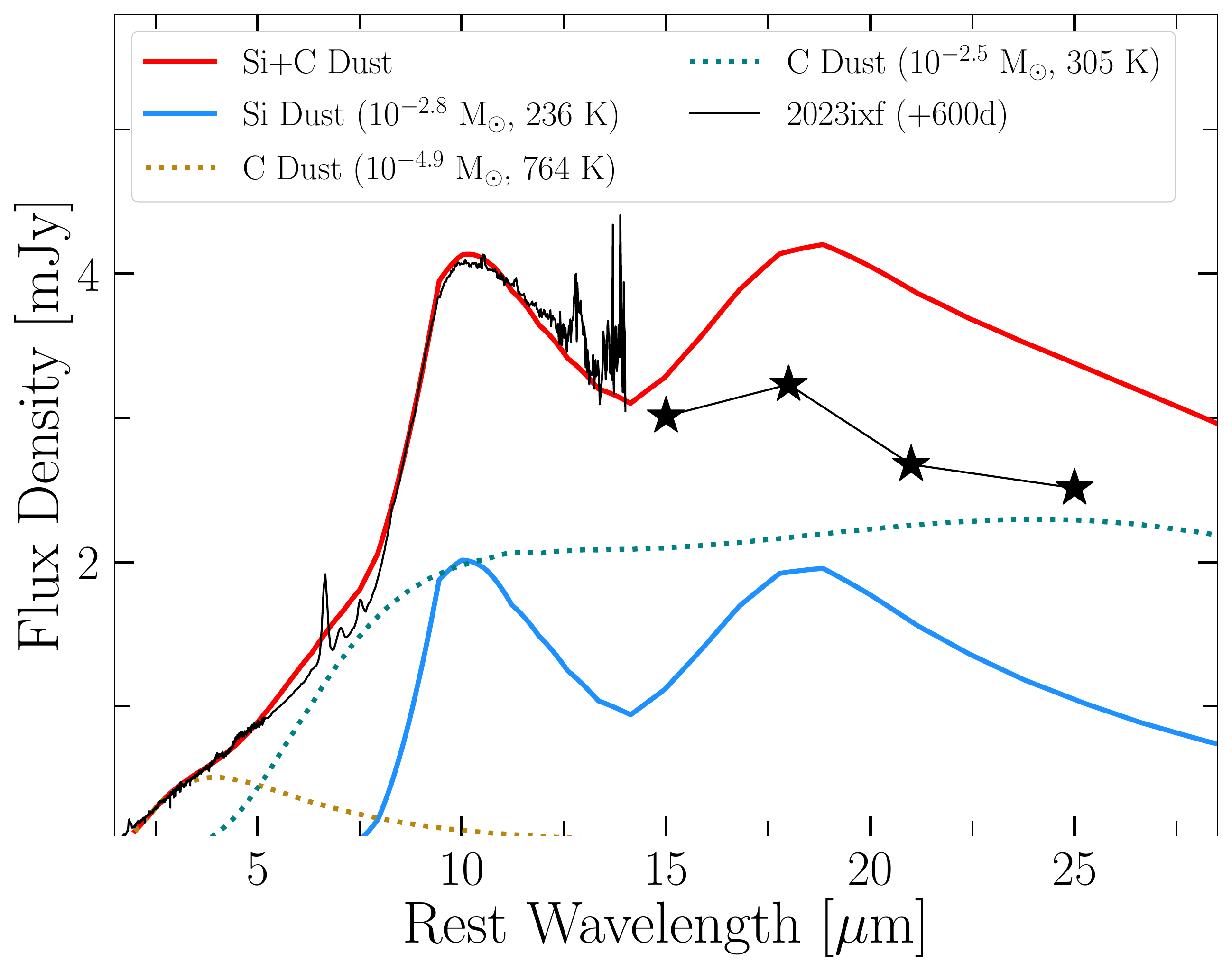}}
\caption{Best-fit, three-component (Si+C dust) dust models using standard analytic prescriptions for optically thin dust (e.g., \citealt{fox11, Tinyanont19, Shahbandeh23, Pearson25}) with opacities from \cite{Draine07}. \label{fig:dust} }
\end{figure*}

\begin{figure*}
\centering
\subfigure{\includegraphics[width=0.99\textwidth]{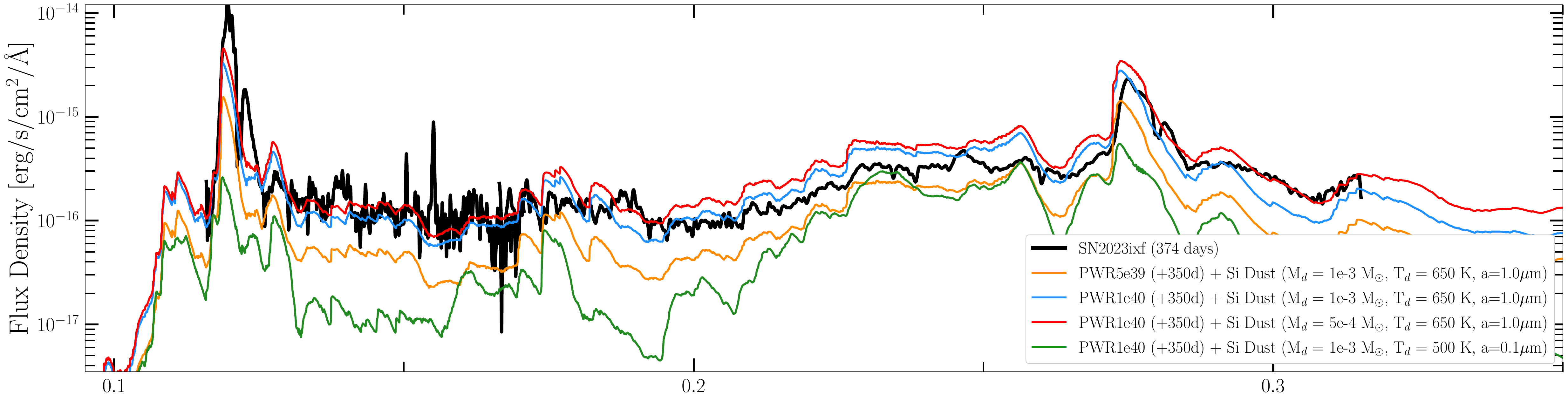}}\\
\subfigure{\includegraphics[width=0.99\textwidth]{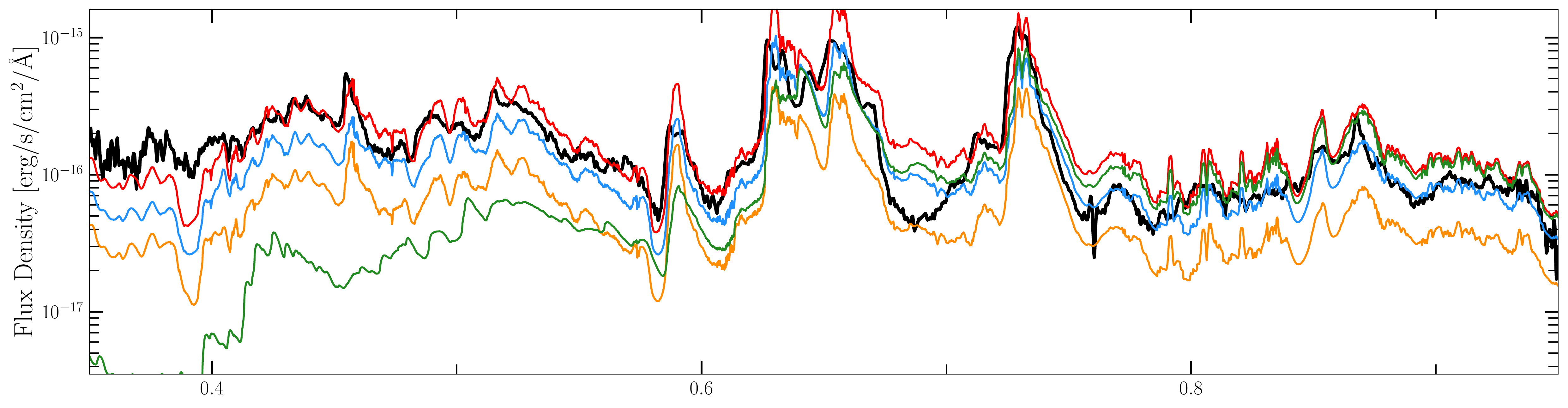}}\\
\subfigure{\includegraphics[width=0.99\textwidth]{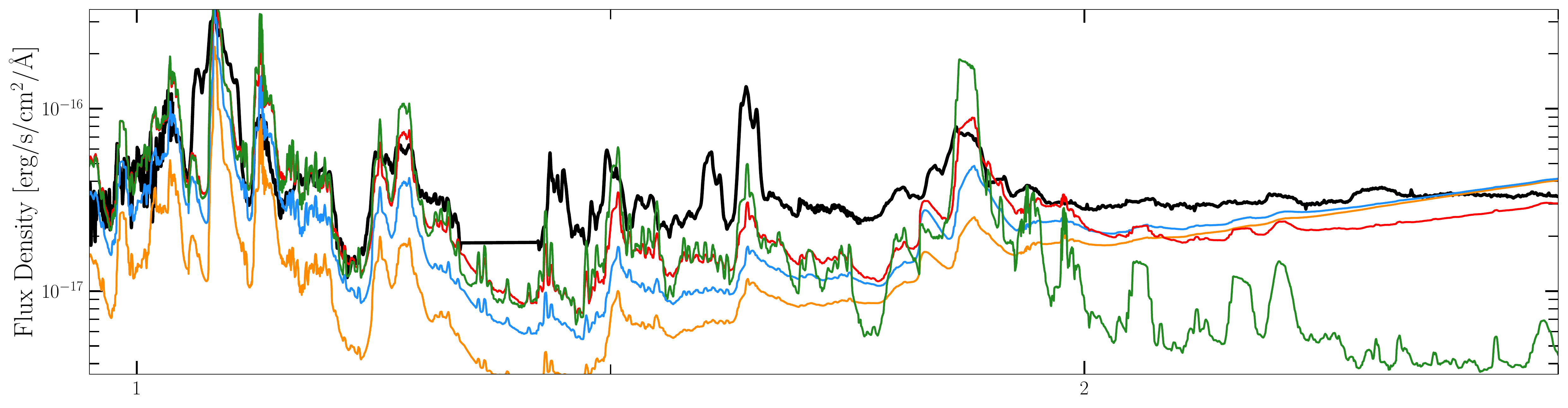}}\\
\subfigure{\includegraphics[width=0.99\textwidth]{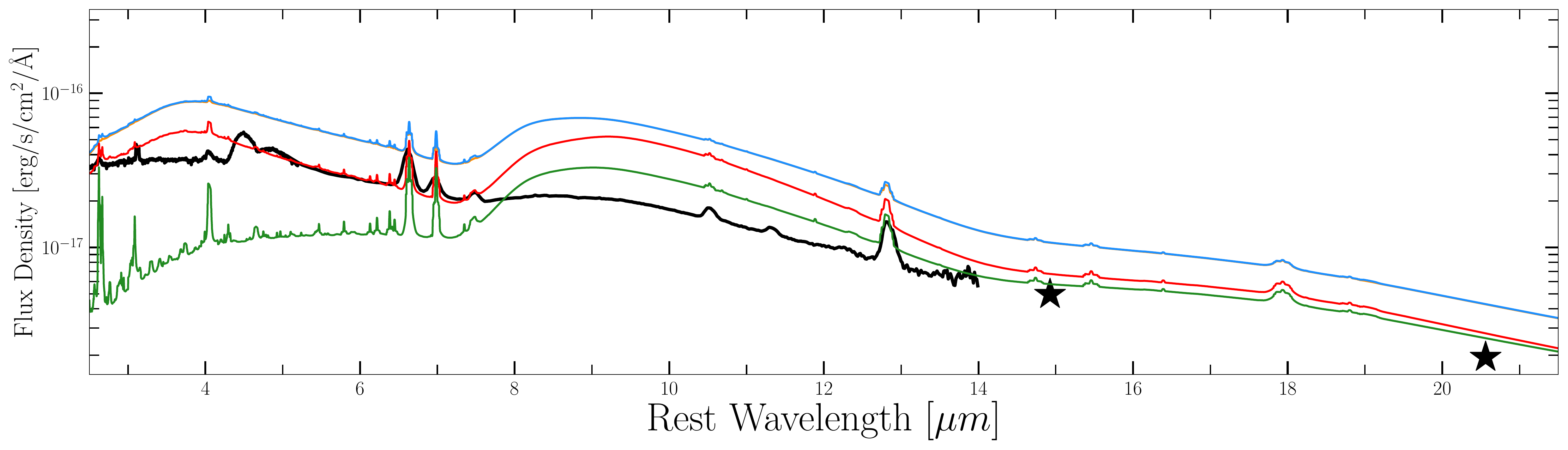}}
\caption{ Zoomed-in comparison of SN~2023ixf UVOIR spectra at $\delta t = 374$~days with respect to {\tt CMFGEN} model spectra at $\delta t = 350$~days for a variety of shock powers and dust parameters (e.g., masses, grain sizes, temperatures). \label{fig:cmfgenall} }
\end{figure*}

Here we present a log of optical spectroscopic observations of SN~2023ixf in Table \ref{tab:spec_all}. In Table \ref{tab:XRT} we present X-ray modeling parameters for single temperature and multi-temperature model fits. Modeling parameters include the temperatures of FS and RS components, intrinsic neutral column densities (fixed for the RS and fit for the FS), and normalizations. For each model fit, we present absorbed and unabsorbed 0.3 - 10~keV fluxes. Figure \ref{fig:XRT} presents {\it Swift}-XRT spectra and best-fit thermal bremsstralung model at three late-time epochs. Figure \ref{fig:dust} shows best-fit dust models to the late-time IR spectra. Figure \ref{fig:cmfgenall} presents variations of \cmfgen\ model spectra at +350~days in order to show the effects of different parameter combinations. 

\counterwithin{figure}{section}

\renewcommand\thetable{A\arabic{table}} 
\setcounter{table}{0}

\section{Additional X-ray Analysis} \label{sec:xrayextra}

HEASoft v6.35.1 was used to extract the late-time {\it Swift}-XRT spectra. Using {\tt xselect v2.5c}, we combined observations carried out roughly within a span of two weeks. A 20'' source region, and a similar size background region positioned to the northeast of the source, were used to extract source and background spectra using {\tt xselect}. The small size of the source region was required to exclude any extraneous X-ray sources. Exposure maps were combined using {\tt ximage} version 4.5.1. The {\tt xrtmkarf } command was used to create the {\it .arf} file, while the {\it swxpc0to12s620210101v016.rmf} file was downloaded as the common {\it .rmf} calibration file for each observation.  The {\tt grppha} command was then used to prepare the spectra for further analysis. Analysis and fitting of the spectra was done using {\tt Sherpa} 4.17.0. The minimum column density was set to the Galactic column density obtained from the HI4PI Survey conducted in 2016 ($7.67 \times 10^{20}$ ~cm$^{-2}$). Given the low statistics, we did not group the spectra, but fit both the source and background spectra simultaneously. The Cash statistic was used as the fitting statistic. This is a better statistic for low-counts data where the traditional chi-squared statistic may be unreliable. The {\tt tbabs} absorption model and {\tt apec} thermal emission model ({\tt xstbabs.abs1*xsvapec.v1}) were used to fit the source spectra, while a power law model matched the background spectra reasonably well. In our {\tt tbabs} model we used the {\tt wilm} set of abundances \citep{Wilms00}. 

\begin{deluxetable*}{ccccccc}
\tablecaption{Ultraviolet/Optical/Infrared Spectroscopy \label{tab:spec_all}}
\tablecolumns{7}
%\tablenum{2}
\tablewidth{0.45\textwidth}
\tablehead{\colhead{UT Date} & \colhead{MJD} &
\colhead{Phase\tablenotemark{a}} &
\colhead{Telescope} & \colhead{Instrument} & \colhead{Wavelength Range} & \colhead{Data Source}\\
\colhead{} & \colhead{} & \colhead{(days)} & \colhead{} & \colhead{} & \colhead{($\mu$m)} & \colhead{}
}
\startdata
2023-09-06T03:34:00 & 60193.1 & 110.3 & Shane & Kast & 0.33-1.1 & YSE \\
2023-09-26T02:57:57 & 60213.1 & 130.3 & Shane & Kast & 0.33-1.1 & YSE \\
2023-11-22T13:31:17 & 60270.6 & 187.7 & Shane & Kast & 0.33-1.1 & YSE \\
2024-02-14T16:11:48 & 60354.7 & 271.8 & Keck & LRIS & 0.32-1.0 & YSE \\
2024-03-19T11:38:19 & 60388.5 & 305.7 & Shane & Kast & 0.33-1.1 & YSE \\
2024-04-12T08:59:13 & 60412.4 & 329.5 & Shane & Kast & 0.33-1.1 & YSE \\
2024-04-18T08:38:13 & 60418.4 & 335.5 & Shane & Kast & 0.33-1.1 & YSE \\
2024-04-29T11:10:42 & 60429.5 & 346.63 & Shane & Kast & 0.33-1.1 & YSE \\
2024-05-17T07:32:49 & 60447.3 & 364.5 & Shane & Kast & 0.33-1.1 & YSE \\
2024-05-29T07:54:55 & 60459.3 & 376.5 & Shane & Kast & 0.33-1.1 & YSE \\
2025-03-04T14:20:32 & 60738.6 & 655.8 & Keck & LRIS & 0.32-1.0 & YSE \\
2023-08-10T05:15:42 & 60166.2 & 83.5 & Shane & Kast & 0.36-1.1 & TReX$^b$ \\
2023-08-10T05:30:49 & 60166.2 & 83.5 & Shane & Kast & 0.58-0.74 & TReX$^b$ \\
2023-08-25T04:14:18 & 60181.2 & 98.4 & Shane & Kast & 0.36-1.1 & TReX$^b$ \\
2023-09-07T03:08:51 & 60194.1 & 111.4 & Shane & Kast & 0.36-1.1 & TReX$^b$ \\
2023-12-05T12:54:31 & 60283.5 & 200.8 & Shane & Kast & 0.36-1.1 & TReX$^b$ \\
2023-12-05T13:12:42 & 60283.6 & 200.9 & Shane & Kast & 0.58-0.74 & TReX$^b$ \\
2024-02-14T10:24:55 & 60354.4 & 271.6 & Shane & Kast & 0.36-1.1 & TReX$^b$ \\
2024-03-18T11:52:24  & 60387.5 & 304.7 & Shane & Kast & 0.36-1.1 & TReX$^b$ \\
2024-03-18T12:40:35 & 60387.5 & 304.7 & Shane & Kast & 0.58-0.74 & TReX$^b$ \\
2024-06-14T06:52:20 & 60475.3 & 392.5 & Shane & Kast & 0.36-1.1 & TReX$^b$ \\
2024-06-14T07:07:56 & 60475.3 & 392.5 & Shane & Kast & 0.58-0.74 & TReX$^b$ \\
2024-08-28T04:06:49 & 60549.2 & 467.4 & Shane & Kast & 0.36-1.1 & TReX$^b$ \\
2024-12-31T15:44:00 & 60675.7 & 592.8 & Keck & KCWI & 0.36-0.89 & Caltech \\
2025-01-26T15:15:56 & 60701.6 & 618.8 & Keck & LRIS & 0.31-1.0 & Caltech \\
2025-04-25T12:42:44 & 60790.5 & 707.7 & Keck & LRIS & 0.31-1.0 & Caltech \\
2024-02-02T15:08:22 & 60342.6 & 259.8 & Keck & NIRES & 0.94-2.5 & KITS \citep{KITS} \\
2024-02-18T14:56:47 & 60358.6 & 275.8 & Keck & NIRES & 0.94-2.5 & KITS \citep{KITS} \\
2024-03-01T14:32:38 & 60370.6 & 287.8 & Keck & NIRES & 0.94-2.5 & KITS \citep{KITS} \\
2024-03-23T14:52:55 & 60392.6 & 309.8 & Keck & NIRES & 0.97-2.5 & KITS \citep{KITS} \\
2024-03-29T13:43:23 & 60398.6 & 315.7 & Keck & NIRES & 0.97-2.5 & KITS \citep{KITS} \\
2024-04-26T13:08:20 & 60426.5 & 347.7 & Keck & NIRES & 0.97-2.5 & KITS \citep{KITS} \\
2024-05-22T11:14:53 & 60452.5 & 369.6 & Keck & NIRES & 1.0-2.5 & KITS \citep{KITS} \\
2025-03-07T15:14:16 & 60741.6 & 658.8 & Keck & NIRES & 0.94-2.5 & KITS \citep{KITS} \\
2025-04-12T12:27:10 & 60776.5 & 694.7 & Keck & NIRES & 0.96-2.5 & Caltech \\
2024-03-24T14:05:30 & 60393.6 & 310.7 & HST & STIS/FUV MAMA & 0.11-0.17 & \citep{ValentiHST} \\
2024-03-24T10:55:46 & 60393.5 & 310.6 & HST & STIS/NUV MAMA & 0.16-0.32 & \citep{ValentiHST} \\
2025-01-26T17:10:18 & 60393.6 & 619 & HST & STIS/FUV MAMA & 0.11-0.17 & \citep{BostroemHST} \\
2025-01-26T14:27:10 & 60393.5 & 619 & HST & STIS/NUV MAMA & 0.16-0.32 & \citep{BostroemHST} \\
2024-01-26T14:34:40 & 60335.6 & 252.8 & JWST & NIRSpec/F170LP & 1.7–3.2 & \citep{Ashalljwst1, Ashalljwst2, Medler25} \\
2024-01-26T14:25:02 & 60335.6 & 252.8 & JWST & NIRSpec/F295LP & 2.9–5.3 & \citep{Ashalljwst1, Ashalljwst2, Medler25} \\
2024-01-26T15:16:11 & 60335.6 & 252.8 & JWST & MIRI/LRS & 5–14 & \citep{Ashalljwst1, Ashalljwst2, Medler25} \\
2024-05-26T13:32:05 & 60456.6 & 373.8 & JWST & NIRSpec/F170LP & 1.7–3.2 & \citep{Ashalljwst1, Ashalljwst2, Medler25} \\
2024-05-26T13:17:58 & 60456.6 & 373.8 & JWST & NIRSpec/F295LP & 2.9–5.3 & \citep{Ashalljwst1, Ashalljwst2, Medler25} \\
2024-05-26T14:07:53 & 60456.6 & 373.8 & JWST & MIRI/LRS & 5–14 & \citep{Ashalljwst1, Ashalljwst2, Medler25} \\
2025-01-08T10:22:19 & 60683.5 & 600.6 & JWST & NIRSpec/F170LP & 1.7–3.2 & \citep{Ashalljwst3, Medler25} \\
2025-01-08T10:09:45 & 60683.5 & 600.6 & JWST & NIRSpec/F295LP & 2.9–5.3 & \citep{Ashalljwst3, Medler25} \\
2025-01-08T11:02:46 & 60683.5 & 600.6 & JWST & MIRI/LRS & 5–14 & \citep{Ashalljwst3, Medler25} \\
\enddata
\tablenotetext{a}{Relative to first light.}
\tablenotetext{b}{TRansient EXtragalactic team at UC Berkeley (PIs Margutti and Chornock)}
%\tablecomments{}
\end{deluxetable*}

\begin{deluxetable*}{cccccccccc}
\tablecaption{X-ray Modeling Parameters \label{tab:XRT}}
\tablecolumns{10}
%\tablenum{2}
\tablewidth{0.45\textwidth}
\tablehead{\colhead{Phase$^a$} &
\colhead{$T_{\rm FS}$} &
\colhead{Norm$_{\rm FS}$} & \colhead{$N_{\rm H,FS}$} & \colhead{$T_{\rm RS}$} & \colhead{Norm$_{\rm RS}$} & \colhead{$N_{\rm H,RS}$} & \colhead{log(Abs. Flux)} & \colhead{log(Unabs. Flux)} & \colhead{BIC}\\
\colhead{(days)} & \colhead{(keV)} & \colhead{($10^{-5}$)} & \colhead{($10^{22}$~cm$^{-2}$)} & \colhead{(keV)} & \colhead{} & \colhead{($10^{22}$~cm$^{-2}$)} & \colhead{(erg s$^{-1}$ cm$^{-2}$)} & \colhead{(erg s$^{-1}$ cm$^{-2}$)} & \colhead{}}
\startdata
$186.2 - 200.4$ & 11.1$^b$ & $2.47_{-0.13}^{+\infty}$ & -- & 0.137$^b$ & $\sim 0$ & 5 & $-12.90_{-0.84}^{+0.12}$ & $-12.86_{-0.84}^{+0.12}$ & 4.7 \\
$186.2 - 200.4$ & 11.1$^b$ & $2.25_{-0.074}^{+1.67}$ & -- & -- & -- & -- & $-12.94_{-0.04}^{+0.17}$ & $-12.91_{-0.04}^{+0.17}$ & 0.84 \\
$499.2 - 524.5$ & 7.0$^b$ & $3.35_{-0.06}^{+2.76}$ & -- & 0.087$^b$ & $0.15_{-0.15}^{+84.4}$ & 2 & $-12.81_{-0.04}^{+0.15}$ & $-11.54_{-0.04}^{+0.15}$ & 3.6 \\
$499.2 - 524.5$ & 7.0$^b$ & $3.32_{-0.04}^{+1.83}$ & -- & -- & -- & -- & $-12.81_{-0.04}^{+0.12}$ & $-12.77_{-0.05}^{+0.11}$ & 0.26 \\
$584.3 - 585.2$ & 6.5$^b$ & $3.63_{-0.15}^{+4.02}$ & -- & 0.087$^b$ & $121.7_{-116.1}^{+793.54}$ & 1.7 & $-12.76_{-0.04}^{+0.16}$ & $-8.81_{-0.04}^{+0.15}$ & 4.5 \\
$584.3 - 585.2$ & 6.5$^b$ & $3.36_{-0.07}^{+4.69}$ & -- & -- & -- & -- & $-12.83_{-0.07}^{+0.22}$ & $-12.78_{-0.07}^{+0.21}$ & 0.70 \\
\hline
$186.2 - 200.4$ & $17.7^{+50.7}_{-12.5}$ & $8.52^{+8.76}_{-2.15}$ & $0.077^{+2.34}_{-0.0}$ & -- & -- & -- & $-13.04_{-0.087}^{+0.16}$ & $-12.85_{-0.12}^{+0.12}$ & -- \\
$499.2 - 514.7$ & $1.44^{+3.20}_{-0.84}$ & $13.9^{+11.6}_{-6.47}$ & $0.089^{+1.83}_{-0.012}$ & -- & -- & -- & $-13.23_{-0.082}^{+0.46}$ & $-12.69_{-0.21}^{+0.20}$ & -- \\
$523.4 - 524.5$ & $2.66^{+11.9}_{-1.87}$ & $12.9^{+73.5}_{-4.12}$ & $0.33^{+2.34}_{-0.25}$ & -- & -- & -- & $-13.11_{-0.097}^{+0.32}$ & $-12.72_{-0.14}^{+0.17}$ & -- \\
$584.3 - 585.2$ & $4.10^{+11.9}_{-1.74}$ & $9.99^{+2.40}_{-2.43}$ & $0.077^{+0.13}_{-0.0}$ & -- & -- & -- & $-12.91_{-0.13}^{+0.10}$ & $-12.81_{-0.12}^{+0.12}$ & -- \\
$646.30$ & $19.5^{+48.9}_{-17.1}$ & $8.04^{+6.26}_{-3.20}$ & $0.41^{+1.43}_{-0.38}$ & -- & -- & -- & $-12.96_{-0.14}^{+0.17}$ & $-12.86_{-0.18}^{+0.15}$ & -- \\
\enddata
\tablenotetext{a}{Relative to first light.}
\tablenotetext{b}{Fixed value in model fit.}
\end{deluxetable*}

\end{document}